\documentclass[sigconf]{acmart}
\AtBeginDocument{%
  }

\setcopyright{acmlicensed}
\copyrightyear{2018}
\acmYear{2018}
\acmDOI{XXXXXXX.XXXXXXX}

\acmConference[CCS ’24]{In Proceedings of the 2024 ACM SIGSAC Conference on Computer and Communications Security}{October 14-18, 2024}{Salt Lake City, U.S.A.}
\acmISBN{978-1-4503-XXXX-X/18/06}

\usepackage{rotating}
\usepackage{graphics}
\usepackage{makecell}
\usepackage{tabularx}
\usepackage{multirow} 
\usepackage{subfigure}
\usepackage{amsmath}
\usepackage{array}
\usepackage{wasysym}
\usepackage{threeparttable}
\usepackage{bm}
\usepackage{algorithm}  
\usepackage{algorithmicx}  
\usepackage{algpseudocode}

\begin{document}

\title{No Vandalism: Privacy-Preserving and Byzantine-Robust Federated Learning}



\author{Zhibo Xing}
\affiliation{%
  \institution{Beijing Institute of Technology}
  \city{Beijing}
  \country{China}}
\affiliation{%
  \institution{University of Auckland}
  \city{Auckland}
  \country{New Zealand}}
\email{3120215670@bit.edu.cn}

\author{Zijian Zhang}
\affiliation{%
  \institution{Beijing Institute of Technology}
  \city{Beijing}
  \country{China}}
\email{zhangzijian@bit.edu.cn}

\author{Zi'ang Zhang}
\affiliation{%
  \institution{Beijing Institute of Technology}
  \city{Beijing}
  \country{China}}
\email{3220231794@bit.edu.cn}

\author{Jiamou Liu}
\affiliation{%
  \institution{University of Auckland}
  \city{Auckland}
  \country{New Zealand}}
\email{jiamou.liu@auckland.ac.nz}

\author{Liehuang Zhu}
\affiliation{%
  \institution{Beijing Institute of Technology}
  \city{Beijing}
  \country{China}}
\email{liehuangz@bit.edu.cn}

\author{Giovanni Russello}
\affiliation{%
  \institution{University of Auckland}
  \city{Auckland}
  \country{New Zealand}}
\email{g.russello@auckland.ac.nz}


\renewcommand{\shortauthors}{Anonymous Submission}

\begin{abstract}
Federated learning allows several clients to train one machine learning model jointly without sharing private data, providing privacy protection. 
However, traditional federated learning is vulnerable to poisoning attacks, which can not only decrease the model performance, but also implant malicious backdoors.
In addition, direct submission of local model parameters can also lead to the privacy leakage of the training dataset. 
In this paper, we aim to build a privacy-preserving and Byzantine-robust federated learning scheme to provide an environment with no vandalism (NoV) against attacks from malicious participants.
Specifically, we construct a model filter for poisoned local models, protecting the global model from data and model poisoning attacks. 
This model filter combines zero-knowledge proofs to provide further privacy protection. 
Then, we adopt secret sharing to provide verifiable secure aggregation, removing malicious clients that disrupting the aggregation process. 
Our formal analysis proves that NoV can protect data privacy and weed out Byzantine attackers. 
Our experiments illustrate that NoV can effectively address data and model poisoning attacks, including PGD, and outperforms other related schemes. 
\end{abstract}




\maketitle

\section{Introduction}
Federated learning is an emerging machine learning approach for data privacy protection that generally consists of a central server and several clients. 
The server aggregates and distributes global models. 
The clients train the global model with their private datasets to obtain local models. 
By aggregating the local models into a new global model, the local datasets can contribute to the global model update, while avoiding direct data disclosure, thus protecting data privacy. 
The emphasis on data privacy under GDPR has boosted the utilization of privacy-preserving federated learning in a variety of applications, including healthcare~\cite{adnan2022federated}, finance~\cite{imteaj2022leveraging}, autonomous driving~\cite{pokhrel2020decentralized}, mobile edge computing~\cite{yu2021toward}, etc.

However, there exists some security threats in federated learning. 
The first issue is about the robustness of federated learning. 
Since the local training process is transparent, malicious clients can perform poisoning attack during the training process.
The aggregation involves poisoned local models can lead to the global model performance degradation, e.g., the drop of accuracy or implanted backdoors in the global model.
To resist poisoning attacks, poisoned local models need to be checked and filter out before the aggregation. 
Common model checking strategies include magnitude-based and direction-based. 
By treating models as vectors and comparing their differences in magnitude and direction, poisoned models can be recognized.
Furthermore, malicious clients can also break the federated learning process during the model aggregation, by sending incorrect or inconsistent messages to other clients or server, or refusing to send messages. 
These kinds of attack is recognized as the Byzantine attack. 
The second issue is about the privacy of federated learning. 
Although avoiding directly sharing local datasets largely protects data privacy, some reconstruction attacks and inference attacks are still able to reconstruct the sample used in the training, or infer which sample is used or not, from the submitted local model. 
These attacks greatly threaten the privacy of federated learning. 
To resist such privacy attacks, secure aggregation protocols are employed. 
Secure aggregation protocols can accomplish the global model aggregation without exposing each local model, which avoids privacy attacks due to local model exposure. 
In general, common secure aggregation protocols are based on homomorphic encryption or secret sharing. 

How to provide good defense performance against poisoning attacks, Byzantine attacks and privacy attacks simultaneously has been a crucial research problem for secure federated learning.
On the one hand, it is necessary to check whether the submitted local model is benign or not without direct access, achieving strong poisoning defense without compromising the privacy. 
On the other hand, it is also necessary to aggregate the checked models in a secure way, to avoid Byzantine attacks and privacy leakage during the aggregation process. 
In addition, the federated learning system should work with ordinary assumptions to enhance the practicality of the scheme. 
For example, a common federated learning system should contain a limited number of clients and only one semi-honest server who does not possess any training data. 
Unfortunately, previous works cannot provide a satisfying defense performance against all these three attacks within ordinary assumptions. 
Thus, we propose a privacy-preserving and Byzantine-robust federated learning system with no vandalism (NoV), which can address aforementioned problems. 
Specifically, we design a novel model filter with layer-wise and hybrid strategy to check the local updates, providing stronger defense performance against data and model poisoning attacks including Projection Gradient Descent~\cite{madry2017towards} (PGD) attack.
This model filter does not require to have a clean dataset. 
Following RoFL~\cite{lycklama2023rofl}, we adopt non-interactive zero-knowledge proofs (NIZK) to make this model filter privacy-preserving.
In addition, we design a novel secure aggregation protocol which can provide privacy protection for the local models while wiping out malicious Byzantine attackers with only one semi-honest server.
The main contributions of our work are shown as follows: 
\begin{itemize}
    \item We propose a model filter with hybrid strategies, which is demonstrated by experiments to be effective in defending against data and model poisoning attacks including PGD attack.
    \item We propose a privacy-preserving secure aggregation protocol based on secret sharing, which protects local models from being leaked while detecting Byzantine attackers during the aggregation.
    \item We provide a privacy-preserving and Byzantine-robust federated learning by combining the above scheme with NIZK. 
    \item We conduct experiments with two well-known datasets to compare the defense performance of NoV with existing schemes. The experimental results show that our scheme performs well with affordable computation time.
\end{itemize}

The rest of this paper is organized as follows: 
Section \ref{Rel} presents existing work in areas related to our paper.
Section \ref{Pre} introduces preliminary knowledge.
Section \ref{Pro} describes the system model, threats and design goals.
Section \ref{Sch} presents our scheme, NoV in detail.
Section \ref{Sec} and Section \ref{Exp} analyze the security and performance of our scheme, respectively.
Section \ref{Con} summarizes our work.

\section{Related Work\label{Rel}} 

In this section, we present an overview on existing researches related to privacy-preserving and Byzantine-robust federated learning. 
For the robustness, we mainly consider its ability in defending against poisoning attacks and Byzantine attacks. 

\begin{table}[t]
\centering
\caption{Comparison with related work in Byzantine-robust federated learning.}
\label{relatedtable}
\begin{threeparttable}
\begin{tabular}{c||cccc}
\hline
\thead{Related Works} & \thead{Poisoning\\Defense} & \thead{Client\\Privacy} & \thead{Byzantine\\Defense} & \thead{Ordinary\\Assumption} \\ \hline 
\thead{FLTrust\cite{cao2020fltrust}} & $\LEFTcircle$ & $\Circle$ & N/A & $\Circle$ \\ 
\thead{SecureFL\cite{hao2021efficient}} & $\LEFTcircle$ & $\Circle$ & N/A & $\Circle$ \\ 
\thead{FLDetector~\cite{zhang2022fldetector}} & $\LEFTcircle$ & $\Circle$ & N/A & $\CIRCLE$ \\ 
\thead{CosDefense~\cite{yaldiz2023secure}} & $\LEFTcircle$ & $\Circle$ & N/A & $\CIRCLE$\\ 
\thead{FLAME~\cite{nguyen2022flame}} & $\CIRCLE$ & $\Circle$ & N/A & $\CIRCLE$ \\ 
\thead{PEFL~\cite{liu2021privacy}} & $\LEFTcircle$ & $\CIRCLE$ & $\CIRCLE$ & $\Circle$ \\ 
\thead{PBFL~\cite{miao2022privacy}} & $\LEFTcircle$ & $\CIRCLE$ & $\CIRCLE$ & $\Circle$ \\ 
\thead{ShieldFL~\cite{ma2022shieldfl}} & $\LEFTcircle$ & $\CIRCLE$ & $\CIRCLE$ & $\Circle$ \\ 
\thead{RoFL~\cite{lycklama2023rofl}} & $\LEFTcircle$ & $\CIRCLE$ & $\LEFTcircle$ & $\CIRCLE$ \\
\thead{Our Work} & $\CIRCLE$ & $\CIRCLE$ & $\CIRCLE$ & $\CIRCLE$\\ 
\hline
\end{tabular}
\begin{tablenotes}
    \footnotesize
    \item[1] $\Circle$ denotes the requirement is not covered; 
    \item[2] $\LEFTcircle$ denotes the requirement is partially covered; 
    \item[3] $\CIRCLE$ denotes the requirement is fully covered;
\end{tablenotes}
\end{threeparttable}
\end{table}

\subsection{Byzantine-Robust Federated Learning}

In federated learning, malicious clients can upload poisoned local updates to spoil the global model, resulting in accuracy degradation or backdoor implantation. 
To defend against such poisoning attacks, 
FLTrust~\cite{cao2020fltrust} provides the server with a \textit{server model} trained by itself on a clean dataset. 
The server will compute the \textit{cosine similarity} between the submitted local model and the server model.
This similarity will be used for calculating the aggregation weights of local models. 
Further, all local models are \textit{normalized} to a same magnitude to reduce the influence of malicious models in terms of magnitude.
SecureFL~\cite{hao2021efficient} follows the same model checking strategy as FLTrust and customizes a number of cryptographic components to increase its efficiency.
In FLDetector~\cite{zhang2022fldetector}, the server computes the \textit{predicted models} for each client $i$. 
The average \textit{Euclidean distance} between the prediction models and their submitted local models over several rounds is utilized as a score. 
By clustering scores among clients, the cluster with higher average scores is identified as malicious clients and subsequently removed.
CosDefense~\cite{yaldiz2023secure} computes the \textit{cosine similarity} of the \textit{last layer parameters} between the local updates and the \textit{global model} as a filtering criterion.
FLAME~\cite{nguyen2022flame} computes \textit{cosine distance} between \textit{each pairwise models} and utilizes the DBSCAN algorithm to cluster these models, obtaining the cluster of malicious models.
PEFL~\cite{liu2021privacy} uses the median of local model parameters. 
The server computes the \textit{Pearson correlation coefficient} between each local model and the \textit{median model}.
This coefficient is treated as the aggregation weight to mitigate the influence of malicious models. 
The privacy is guaranteed by homomorphic encryption among two servers. 
PBFL~\cite{miao2022privacy} utilizes a clean dataset and cosine similarity to identify possible malicious updates. With a homomorphic encryption-based aggregation scheme, the local model privacy could be preserved with two non-colluding central server. 
ShieldFL~\cite{ma2022shieldfl} examines the \textit{cosine similarity} between the \textit{normalized} local models and the \textit{global model} from the previous round. 
Homomorphic encryption is utilized to protect data privacy during the computation between the two server. 
RoFL~\cite{lycklama2023rofl} calculates the \textit{$L_2$} norm and \textit{$L_\infty$} norm between the submitted local model and the \textit{global model from the previous round}. 
If the norm exceeds a pre-defined threshold, then that update will be filtered out.

However, existing schemes suffer from following issues: 
\begin{itemize}
    \item Vulnerable to specific attacks: Most schemes~\cite{zhang2022fldetector,ma2022shieldfl,lycklama2023rofl,liu2021privacy,cao2020fltrust,miao2022privacy,hao2021efficient} rely on simple model checking strategies based on Euclidean distance or cosine similarity. 
    However, when facing stronger model poisoning attacks with non-independent identically distributed (non-iid) datasets, these defense is much less effective. 
    PGD~\cite{madry2017towards} attack can project the malicious model into a certain range of the reference model, thus bypassing these defense.
    
    To address stronger poisoning attacks, we propose a hybrid and layer-wise model checking strategy, further identifying the difference between malicious and benign updates.
    \item Breaking local privacy: Some schemes~\cite{cao2020fltrust,zhang2022fldetector,yaldiz2023secure,nguyen2022flame,hao2021efficient} assume that the server has direct access to all local models to compute their magnitude and direction. 
    However, considering existing data reconstruction attacks and membership inference attacks, the direct access to local models will compromise the privacy of local datasets.
    
    To address the privacy leakage during the model checking, we employ NIZK, converting the model checking process into a zero-knowledge one.
    \item Rely on extra prior knowledge: Some schemes~\cite{cao2020fltrust,miao2022privacy,hao2021efficient} assume that the server has a clean dataset, which can be used to train a benign model as the reference model. 
    By comparing with this reference model, malicious models can be identified. 
    This assumption requires the server to have additional computational and data collection capabilities, reducing the application scenarios of the schemes.

    To address the additional reliance on clean dataset, we use the previous global model as the reference model. 
    With a more efficient model filter, malicious models can be filtered out.
\end{itemize}

\subsection{Privacy-Preserving Federated Learning}

In federated learning, clients only need to submit local models, avoiding direct exposure of the local dataset. 
Nevertheless, several reconstruction attacks~\cite{zhu2019deep,zhao2020idlg,geng2021towards} have demonstrated that submitted local models can lead to the leakage of local datasets. 
Therefore, privacy-preserving aggregation methods are necessary to prevent the reveal of local models.
Existing privacy-preserving aggregation schemes, integrated with defense against poisoning, can be categorized into secret sharing-based (SS-based) and homomorphic encryption-based (HE-based). 
For instance, PEFL~\cite{liu2021privacy}, ShieldFL~\cite{ma2022shieldfl}, and PBFL~\cite{miao2022privacy} employ HE-based aggregation schemes, while RoFL~\cite{lycklama2023rofl} utilizes a SS-based aggregation scheme.
HE-based schemes protect the privacy of local update through the homomorphic aggregation on ciphertext. 
However, HE-based schemes always introducing the assumption of several non-colluding server for secure aggregating.
SS-based schemes protect the privacy of local updates by dividing the local update into several shares.
These shares leak no information that contribute to infer local updates, and can be aggregated for reconstruction the global update.
However, it is difficult for SS-based schemes to detect the Byzantine attackers. 
Thus once the attack occurs, the current training round has to be rolled back.
Specifically, existing schemes suffer from following issues: 
\begin{itemize}
    \item Rely on multiple non-collude server: Some schemes~\cite{ma2022shieldfl,liu2021privacy,miao2022privacy} achieve privacy-preserving aggregation through multiple non-colluding semi-honest servers. 
    This assumption reduces the security of the scheme, leading to limited application scenarios. 

    To address the additional reliance on multiple non-colluding semi-honest servers, we adopt a SS-based aggregation protocol.
    \item Vulnerable to Byzantine attacks: The scheme~\cite{lycklama2023rofl} based on secret sharing less considers the malicious behavior during the aggregation, such as sending different messages for different participants, intentionally sending error messages, or refusing to communicate. 
    Such Byzantine attacks can pose a great challenge to aggregation protocols. 
    RoFL can only recognize if an attack has occurred, and do not limit the attackers and their malicious actions.

    To address the vulnerability on Byzantine attacks, we propose a verifiable aggregation protocol that utilizing verifiable secret sharing and verifiable decryption to identify Byzantine attackers.
\end{itemize}


\section{Preliminaries\label{Pre}}

In this section, we primarily introduce some background knowledge relevant to our scheme, including federated learning, poisoning attacks, zero-knowledge proofs, and secret sharing.

\subsection{Federated Learning}

Federated learning (FL)~\cite{mcmahan2017communication} is a distributed machine learning technique that allows multiple clients to jointly train a global model using their respective private datasets. 
In general, a federated learning includes a central server $S$ and $n$ clients $C_1, ..., C_n$, where each client $C_u$ possesses a private dataset $D_u$. 
Taking the $j$-th round of training as an example: 
\begin{enumerate}
    \item Server $S$ distributes the global model $M_{j-1}$ from the previous round to each client. 
    \item Each client $C_u$ train the global model with her private dataset $D_u$ to obtain the local update $\delta^j_u$. 
    \item Each client $C_u$ submits her local update $\delta^j_u$ to the server $S$, with which the server can aggregate the local updates submitted by clients with $M^j = M^{j-1} + \frac{\Sigma_{u=1}^n \delta^j_u}{n}$ to obtain the updated global model.
\end{enumerate}
And this process iterates until the global model converges.
  
\subsection{Poisoning Attacks}

Poisoning attacks aim to degrade the performance of a model. 
In federated learning, adversaries achieve this by submitting poisoning local models, which can eventually impact the performance of the global model. 
Depending on the goal, poisoning attacks can be classified as untargeted attacks and targeted attacks.
The objective of an untargeted attack is to degrade the model's performance across all inputs without focusing on specific target inputs~\cite{fang2020local}. 
On the other hand, a targeted attack aims to influence the model's performance on specific inputs while maintaining the overall performance on the task~\cite{bagdasaryan2020backdoor}. 
For instance, it may focus on misclassifying certain images. 
Based on the method, poisoning attacks can be divided into data poisoning and model poisoning.
Data poisoning involves malicious modifications to the training dataset to induce poison in the trained model~\cite{tolpegin2020data}. 
For example, flipping the labels of training data to make the model learn incorrect classification results. 
Model poisoning, on the other hand, involves directly manipulating the weights of a trained model to adversely impact its performance on the target task~\cite{wang2020attack}.

\subsection{Zero-Knowledge Proof}

Zero-knowledge proof (ZKP)~\cite{goldwasser1989knowledge} is a cryptographic protocol used for a prover $P$ to convince a verifier $V$ that a specific statement is true without revealing any additional information during the process. 
More formally, ZKP is a protocol to prove knowledge of a witness $w$ for a statement $\phi$ that satisfying $(w, \phi)\in R$, a NP relation that defines whether a given witness is valid for the statement or not.
In essence, a zero-knowledge proof protocol has three properties:
\begin{enumerate}
    \item Completeness: If the statement is true, an honest verifier will be convinced by an honest prover.
    \item Soundness: If the statement is false, no cheating prover can convince an honest verifier.
    \item Zero-knowledge: No additional information about the statement, apart from its truth, is revealed to the verifier.
\end{enumerate}
In NoV, we utilize zero-knowledge proof for the implementation of the model filter to ensure the protection of local model privacy.

\subsection{Verifiable Secret Sharing}

Secret sharing (SS) is a cryptographic scheme used to split and reconstruct secrets. 
For the t-out-of-n secret sharing scheme employed, it enables the division of a secret $s$ into $n$ shares. 
The original secret $s$ can be reconstructed using any arbitrary set of $t$ shares, while any set of $t-1$ shares remains insufficient to retrieve any information about $s$. 
Specifically, the algorithm $\mathrm{SS.share}(s, t, \mathcal{U})\rightarrow \{(u, s_u)\}_{u \in \mathcal{U}}$ takes the secret $s$, threshold $t$, and a list of users $\mathcal{U}$ as input, generating shares $s_u$ for each user $u\in \mathcal{U}$. 
The algorithm $\mathrm{SS.recon}(\{(u, s_u)\}_{u \in \mathcal{V}}, t)\rightarrow s$ takes threshold $t$ and shares $s_u$ from $|\mathcal{V}|\geq t$ users as input and reconstructs the original secret $s$. 
Verifiable secret sharing (VSS)~\cite{pedersen1991non} enables users to verify whether a given share corresponds to the same secret. 
Additionally, the secret sharing scheme~\cite{pedersen1991non} used in this paper exhibits additive homomorphism~\cite{benaloh1986secret}, means for secrets $s^1$ and $s^2$, $\mathrm{SS.recon}(\{(u, s^1_u+s^2_u)\}_{u \in \mathcal{V}}, t)\rightarrow s^1+s^2$.
In NoV, we utilize verifiable secret sharing for constructing the aggregation protocol to identify malicious Byzantine attackers.

\subsection{Public-Key Encryption}

Public-key encryption is a cryptographic scheme used to encrypt and decrypt messages for secure communication.
Specifically, the ElGamal encryption on group $\mathbf{G}$ of order $q$ with generator $g$ used in our scheme is described as follow: \\
- KeyGen($1^\lambda$) generates private-public key pairs $(sk, pk)$ where $sk\in_R \mathbf{Z}^*_q$, $pk=g^{sk}$;\\
- Enc($pk, m$) generates ciphertext $c=(pk^r, g^rm)$ of message $m\in \mathbf{G}$ with public key $pk\in \mathbf{G}^*$ where $r\in_R \mathbf{Z}_q$;\\
- Dec($sk, c=(c_1, c_2)$) generates plaintext $m=c_1^{-\frac{1}{sk}}c_2$ of ciphertext $c\in \mathbf{G}^2$ with private key $sk\in \mathbf{Z}^*_q$.
In NoV, we utilize public-key encryption to ensure the security and privacy of communication between clients.

\section{Problem Statement\label{Pro}}

In this section, we describe the system model, the threat model, and the design goals of our scheme, respectively.  

\subsection{System Model}

\begin{figure}
    \centering
    \includegraphics[width=0.49\textwidth]{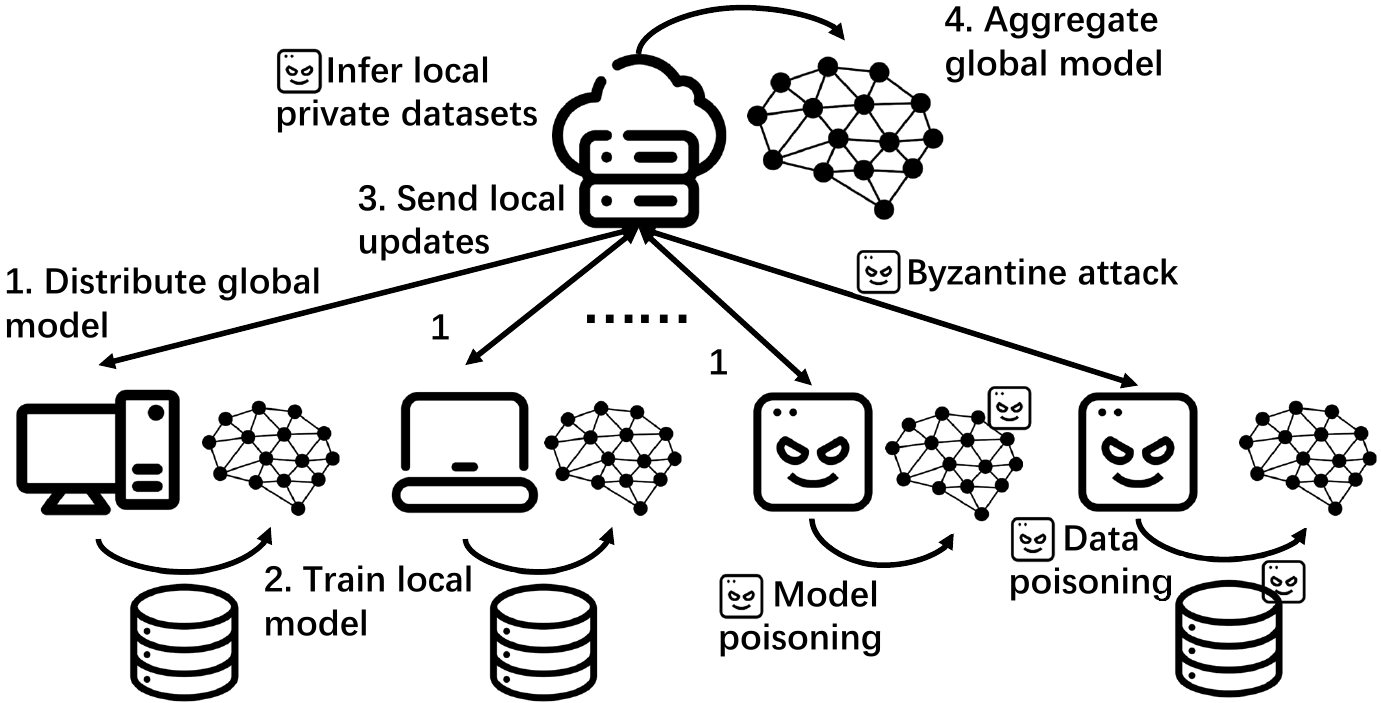}
    \caption{The system and threat model for NoV.} 
    \label{fig/system}
\end{figure}

Figure \ref{fig/system} illustrates the system architecture. 
In this system, there are a total of $n$ clients denoted as $C_1$ to $C_n$ and a server $S$. 
The primary objective of these participants is to collaborate with other, using their respective private dataset to train an optimized model without compromising privacy. 
Each client has a pair of public and private keys $(PK_i, SK_i)$ for encrypted communication and a private dataset $D_i$ for local training.
During the $j$-th round, server $S$ distributes global model $M_{j-1}$ to all clients. 
Client $C_i$ trains global model $M_{j-1}$ with their private dataset $D_i$, resulting in a local model $m^i_j$. 
To guarantee the validity of the local model, client $C_i$ is also required to generate a proof $\pi^i_j$ for their local model $m^i_j$.
By sending the proofs and local updates, clients can participate in the aggregation of global model. 
Through a secure aggregation protocol, $S$ could obtain the updated global model $M_j$ aggregated from valid local updates $\{m^i_j\}_{valid}$.

\subsection{Threat Model}

Figure \ref{fig/system} illustrates the security threats to the system. 
In this system, the server is assumed to be curious, which means that the server will faithfully execute the protocol, but will try to infer local private updates or datasets from clients. 
If the curious server has access to one local model, then by using deep leakage attacks~\cite{zhu2019deep,zhao2020idlg,geng2021towards} or member inference attacks~\cite{zhang2020gan,hu2021source,wang2022poisoning}, the privacy of the corresponding training dataset will be destroyed.
Besides, it is assumed that the server will not collude with any clients.

All the clients are assumed to be curious, they also will try to infer private datasets from others. 
Further, for a system contains $n$ clients, this scheme can tolerate $t-1$ colluding malicious clients, where $t$ represents the minimum number of participants required in a $(t, n)$ threshold secret sharing scheme to reconstruct a secret. 
Malicious clients may further attempt various types of attacks, including:
\begin{itemize}
    \item Poisoning attacks: These attacks aiming at injecting backdoors~\cite{bagdasaryan2020backdoor,wang2020attack} or decreasing the accuracy~\cite{fang2020local} of the global model.
    \item Byzantine attacks: Byzantine attackers try to disrupt the federated learning process by intentionally submitting incorrect local models or refusing to submit local models.
    \item Collusion attacks: Malicious clients can collude with each other to attempt infer local models and datasets from other clients.
\end{itemize}

\subsection{Design Goals}
Our goal is to design a privacy-preserving and Byzantine-robust federated learning system. 
Given the threat model above, the design goals for this system are as follows:
\begin{itemize}
    \item Robustness: The federated learning system should be resilient to poisoning attacks and Byzantine attacks from malicious clients. 
    The aggregation procedure and the accuracy of the global model should not be affected by malicious clients.
    \item Privacy: The federated learning system should protect against the curious server and curious or malicious clients attempting to access or infer the local model or the dataset from other curious clients. 
    No other participants should be able to access the parameters in local models from a curious clients.
\end{itemize}

\section{The NoV Scheme\label{Sch}}

In this section, we first provide a technical overview. 
Then we present the specific construction of two main components in NoV, model filter and aggregation protocol.

\subsection{Technical Overview}

\begin{figure}
    \centering
    \subfigure[Difference on magnitude and direction distribution.]{
    \includegraphics[width=0.22\textwidth]{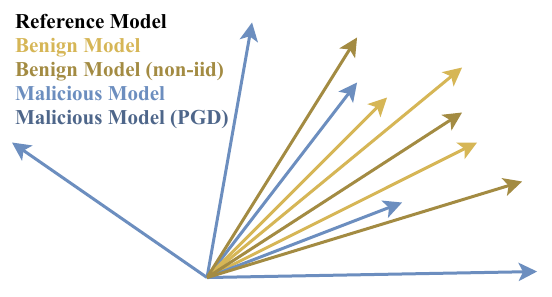}
    \label{over_1}
    }
    \subfigure[PGD attack bypasses model checking strategies.]{
    \includegraphics[width=0.22\textwidth]{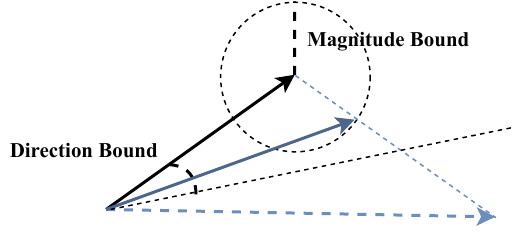}
    \label{over_2}
    }
    \subfigure[Impact of poisoning attack on different layers.]{
    \includegraphics[width=0.25\textwidth]{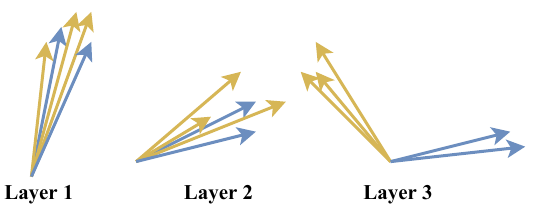}
    \label{over_3}
    }
    \caption{Observations on malicious models and benign models.}
    \label{overview}
\end{figure}

A common observation~\cite{cao2020fltrust} is that benign models follow a similar distribution of magnitude and direction, but malicious models not. 
Thus malicious models can be identified by comparisons in magnitude or direction. 
Besides, non-iid data among clients can lead to a wider distribution of benign models, making it more difficult to identify malicious models. 
This observation is depicted in figure \ref{over_1} and further supported in figure \ref{dis_exp}. 
Another observation~\cite{yaldiz2023secure} is that data poisoning attacks do not significantly affect the parameters of the entire model, but only a small portion of the layers, as depicted in figure \ref{over_3}.
When the model size increases, the poisoned model will be more difficult to be recognized, as the malicious CNN models are more obvious than the LeNet5s in figure \ref{dis_exp}. 
Besides, when the range of samples affected by the data poisoning attack decreases, the impact on the model also decreases. 
Thus, backdoor attacks are more difficult to recognize compared to label flipping attacks, and backdoor attacks using tail data are also more difficult to recognize than general backdoor attacks, which is also shown in figure \ref{dis_exp}. 
Tail data refers to special samples with small amount under a certain classification, such as number 7 with a horizontal bar, or a specific model of airplanes. 
In addition, PGD attack can project malicious models into a specific range of reference model with a small loss of attack efficiency. 
As a result, most magnitude and direction-based model checking strategies cannot provide satisfying defense performance against PGD attack, as depicted in figure \ref{over_2}. 
Therefore, to enhance the defense performance of the model filter, we propose a hybrid and layer-wise model checking strategy. 
First of all, we check local model updates rather than local models itself, because the local model contains the global model of the previous round, which can dilute the impact of poisoned local updates. 
In the model filter, we first constrain the local updates in magnitude to exclude local models that are significantly far from the reference model. 
Then we impose a layer-wise direction constraints on the local updates. 
Layer-wise strategies not only improves the model filter's ability to detect poisoning attacks, but also minimize the impact of increasing model size. 

For the aggregation protocol, a technical route based on secret sharing was chosen in order to eliminate the reliance on multiple non-colluding semi-honest servers. 
However, this leads to that malicious clients can submit inconsistent shares to corrupt the aggregation result. 
Although the server can detect the vandalism after aggregation, most schemes, including RoFL, choose to abort the protocol and start a new training round with different clients. 
However, this is not realistic for scenarios with a small number of clients. 
Therefore, by combining verifiable secret sharing with verifiable decryption, we propose a verifiable aggregation protocol. 
By verifying the correctness of the shares as well as the decryption result, the server can identify and kick out the specific malicious Byzantine attacker and resume the aggregation result for this round.

\subsection{Model Filter\label{section_mc}}


Based on the aforementioned observations, we propose a model filter with hybrid and layer-wise strategy for the server filtering out malicious local model updates.
Each client $C_u$ is required to provide a proof with their submitted local update $\delta$, arguing its benignity, to pass the model filter. 
The hybrid model filter strategy involves both magnitude and direction constraints on predefined threshold $t_m$ and threshold $t_d$. 
Further, we adopt layer-wise direction constraints to improve the performance of the model filter on larger machine learning models.
Specifically, we adopt the Euclidean distance and cosine similarity:
\begin{equation}
    \label{equation_constraint1}
    ||\delta|| \leq t_m
\end{equation}
\begin{equation}
    \label{equation_constraint2}
    \frac{\delta_l\cdot M_l}{||\delta_l||\cdot||M_l||} \geq t_d
    ,\ 
    l\in [1,L]
\end{equation}
Where $M$ represents the reference model, which is the global model from previous round.
$L$ represents the number of layers of model $M$, and $M_l$ denotes the vector comprising the parameters in $l$-th layer of model $M$.
Equation \ref{equation_constraint1} limits the magnitude of the update to threshold $t_m$ to avoid malicious updates beyond that range from affecting the global model. 
Equation \ref{equation_constraint2} provides a more precise filtering on the direction of the update. 
The difference in direction between benign and malicious updates becomes insignificant as the model becomes larger. 
Meanwhile, non-iid datasets also lead to a wider distribution of benign models.
Therefore we conduct a layer-wise strategy to examine the structure of the updates more carefully. 
In our model filter, for the check of equation \ref{equation_constraint1}, those updates that are not satisfied will be discarded. 
For the check of equation \ref{equation_constraint2}, the number of layers that can satisfy the check will be recorded. 
Then all these local updates will be sorted in descending by this number and only the top $n*t_s$ updates will be selected for aggregation, where $t_s$ is the selection percentage.

Specifically, NoV adopts a zero-knowledge range proof scheme, Bulletproofs~\cite{bunz2018bulletproofs}, to generate and verify the proofs for the aforementioned constraints in equation \ref{equation_constraint1} and \ref{equation_constraint2}. 
For equation \ref{equation_constraint1}, considering $\delta$ as a vector $(x_1, ..., x_p)$ of length $p$, the aim is to prove $\Sigma_{i=1}^p {x_i}^2 \leq {t_m}^2$ holds. 
Each $x_i$ and ${x_i}^2$ can be linked through a $\Sigma$-protocol with their respective Pedersen commitment $c_i=Com(x_i)=g^{x_i}h^{r_i}$ and $c'_i=Com(x^2_i)=g^{x^2_i}h^{r'_i}=c_i^{x_i}h^{r'_i-x_i r_i}$.
Exploiting the homomorphic property of commitments, the proof for equation \ref{equation_constraint1} can be completed with commitment $C'=\Pi_{i=1}^p c'_i=g^{\Sigma_{i=1}^p x^2_i}h^r$ and a range proof of $\Sigma_{i=1}^p x^2_i \leq t_m^2$. 

As for equation \ref{equation_constraint2}, we set the similarity threshold $t_d$ as 0, thus the check transformed into $\delta_l\cdot M_l\geq 0$ for each layer. 
Both $\delta_l$ and $M_l$ can be regarded as vector $(x_1, ..., x_q)$ and $(y_1, ..., y_q)$ of length $q$, respectively. 
With the commitment $c_i=Com(x_i)=g^{x_i}h^{r_i}$ and the known $M_l$, it is sufficient to prove equation \ref{equation_constraint2} by proving $\Sigma_{i=1}^q x_i y_i \geq 0$ with commitment $C=\Pi_{i=1}^q c_i^{y_i}=g^{\Sigma_{i=1}^q x_i y_i}h^r$. 

In NoV, the threshold $t_m$ and $t_d$ are assumed to be public, thus both the honest and malicious clients can project their local update into the given range with PGD.

\subsection{Aggregation Protocol}

\begin{figure}[h]
    \centering
    \includegraphics[width=0.49\textwidth]{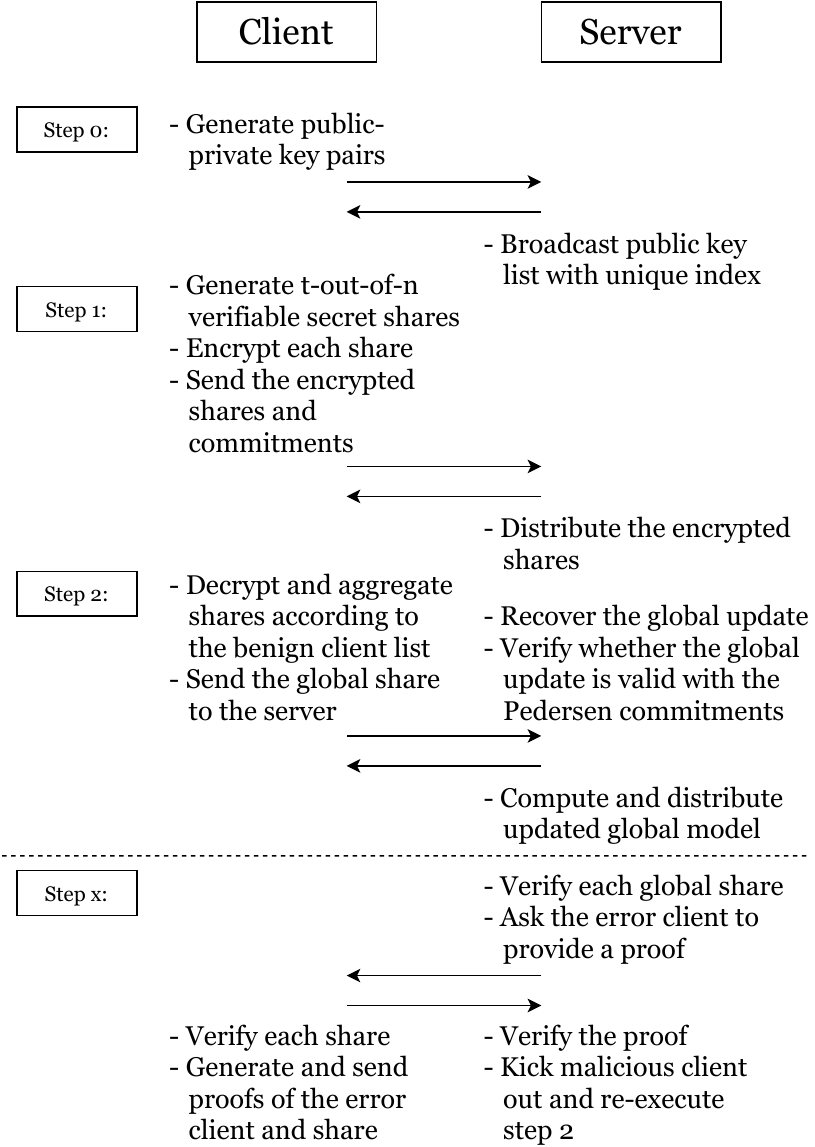}
    \caption{The workflow of the Model aggregation protocol.} 
    \label{fig/aggregation}
\end{figure}

To ensure the local model privacy of clients while guaranteeing the correctness of the aggregation procedure, we propose a verifiable aggregation protocol. 
Specifically, we adopt verifiable secret sharing~\cite{pedersen1991non} and verifiable decryption~\cite{bootle2015short} to provide verifiability for identifying malicious Byzantine attackers.
Their corresponding pseudo-codes of algorithms for generating and reconstructing secret shares, and proving the correctness of decryption, including VSS\_Gen(), Prv\_Dec, VSS\_Rec(), are provided in Appendix \ref{pseudo}. 
Figure \ref{fig/aggregation} illustrates the workflow of the aggregation protocol:



\begin{enumerate}
    \item[0)] Each client $C_u$ generates their public-private key pairs $(PK_u,$ $SK_u)$ with the key generation algorithm $KeyGen(1^\lambda)$ and send their public key $PK_u$ to the server. 
    Server $S$ assigns a unique index $u\in [1,n]$ to each client and broadcasts the list of index along with corresponding public key $\{(u, PK_u)\}_{u\in [1,n]}$ to all clients.
    This initialization step only needs to be performed once.

    \item[1)] For each parameter $x_u$ in the local update of client $C_u$, with a Pedersen commitment $g^{x_u}h^{r_u}$, a verifiable secret sharing scheme~\cite{pedersen1991non} can be employed.
    Running the secret sharing generation algorithm $VSSGen(g, h, x_u, r_u, t, n)$, client $C_u$ can create a $t$-$n$ secret share $\{(s_{u,v}, o_{u,v})\}_{v\in [1,n]}$ for parameter $x_u$.
    In addition, a set of commitments $\{E_{u,j}\}_{j\in [0,t-1]}$ for verifying whether a certain share corresponds to $x_u$ with $r_u$ is also generated.
    Client $C_u$ encrypts each $(s_{u,v}, o_{u,v})$ with the public key $PK_v$ of client $C_v$ and sends all these encrypted shares and commitments to the server.
    Server sends the encrypted shares to each specific client.
    Besides, based on the result of model checking, server can obtain a benign client list $U_B$ whose parameters can be used to aggregate the global model, which will also be broadcasted to all the clients.
    
    \item[2)] From each client $C_v$, client $C_u$ receives and decrypts the ciphertext to obtain share $(s_{v,u}, o_{v,u})$. 
    Then client $C_u$ aggregates all the shares submitted by benign clients according to $U_B$, to obtain $S_u = \Sigma_{v \in U_B} s_{v,u}$ and $O_u = \Sigma_{v \in U_B} o_{v,u}$, a share of one parameter $X$ in the global update, and sends it to the server. 
    With the secret sharing reconstruction algorithm $VSSRec(t, \{(S_u, O_u)\}_{u \in U_B})$, the server can reconstruct the parameter $X=\Sigma_{u\in U_B} x_u$ and $R=\Sigma_{u\in U_B} r_u$ of the global update. 
    It can be easily verified whether the recovered $X$ and $R$ align with the Pedersen commitment with equation \ref{simple_check} holds.
    If this verification passes, server will start the next round training.
    If this verification fails, server will raise an extra step to identify the Byzantine attacker.
    
    \begin{equation}
        \label{simple_check}
        \Pi_{u\in U_B} g^{x_u}h^{r_u} = g^Xh^R
    \end{equation}

    \item[x)] The server can verify whether the share $S_u$ and $O_u$ are aggregated from benign client list through equation \ref{verify_globalshare}, where $E^*_j=\Pi_{u\in U_B}E_{u,j}$ from benign clients. 
    If the verification for client $C_u$ fails, $C_u$ will be asked to provide a $\Sigma$-protocol proof, arguing that at least one share $(s_{v,u}, o_{v,u})$ she received is invalid. 
    Server will send all the commitments generated in step 1 to $C_u$ for the verification.
    $C_u$ can check whether share $(s_{v,u}, o_{v,u})$ she received aligns with commitments $\{E_{v,j}\}_{j\in [0,t-1]}$ given by $C_v$ by checking if equation \ref{verify_share} holds. 
    If the verification for $C_v$ fails, $C_u$ can send a proof $\pi$ following the $\Sigma$-protocol~\cite{bootle2015short} generated by algorithm \ref{prove_decrypt} Prv\_Dec($PK_u, SK_u, (s_{v,u}, o_{v,u}), c$) to the server, arguing that client $C_v$ provides an incorrect share.
    Server will verify whether the decryption is correct by checking equation \ref{verify_decrypt} and whether the share is incorrect by checking equation \ref{verify_share}. 
    If equation \ref{verify_decrypt} holds while equation \ref{verify_share} not, the server will kick client $C_v$ out as a malicious client. 
    Otherwise, client $C_u$ will be kicked.
    The malicious client will be wiped out from the client list, the server can re-broadcasts a benign client list and re-execute step 2.
    
    \begin{equation}
        \label{verify_globalshare}
        g^{S_u}h^{O_u} = \Pi_{j=0}^{t-1} {E_j^*}^{(u^j)}
    \end{equation}
    \begin{equation}
        \label{verify_share}
        g^{s_{v,u}}h^{o_{v,u}}=\Pi_{j=0}^{t-1} {E_{v,j}}^{(u^j)}
    \end{equation}
    \begin{equation}
        \label{verify_decrypt}
        g^e·A = g^z \land c_1^{e}\cdot B=(c_2m^{-1})^z
    \end{equation}

\end{enumerate}

We provide the full aggregation protocol in figure x. 

\begin{figure*}[t]
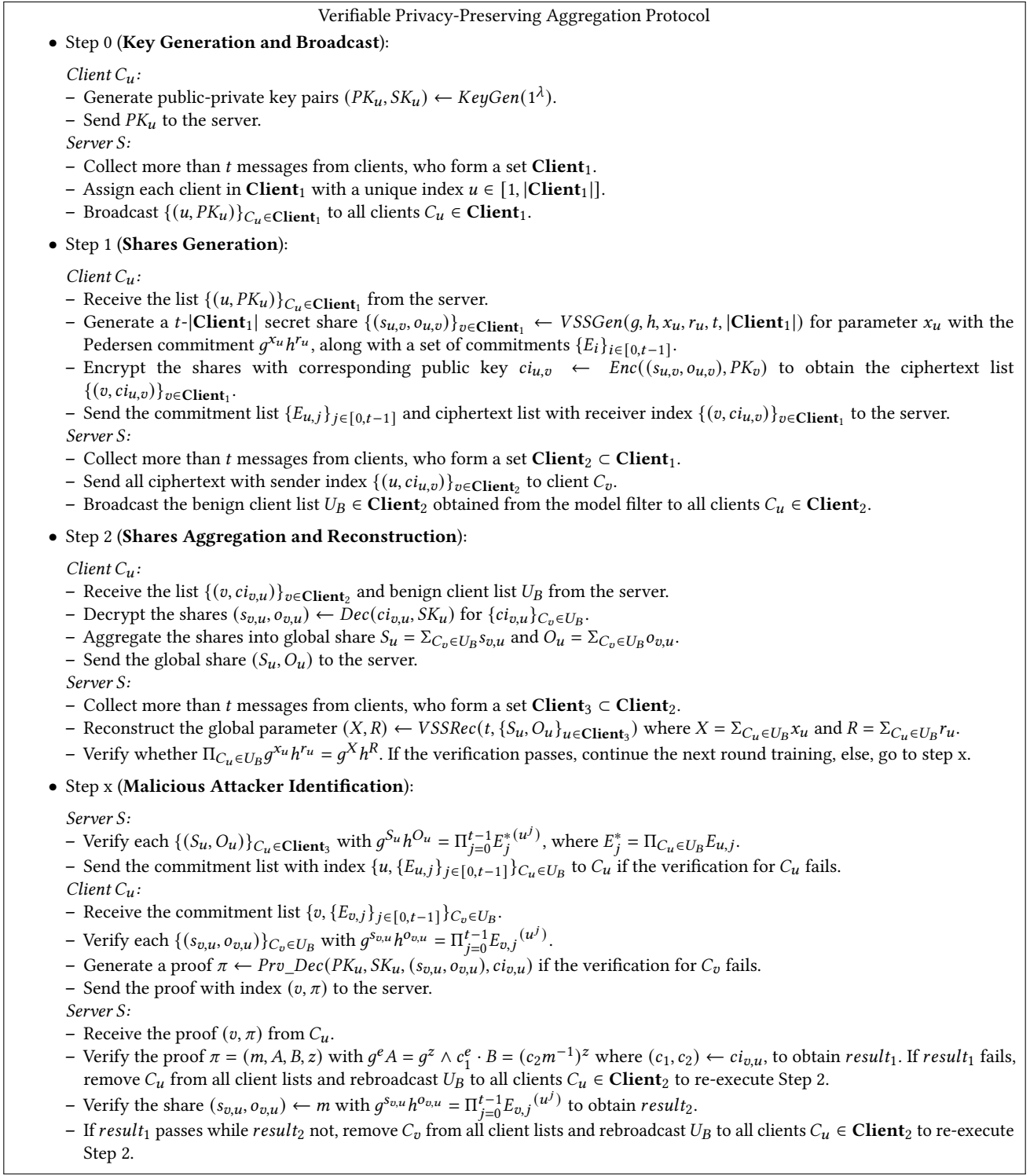

    \centering
    \fbox{
	\parbox{0.95\linewidth}{
            \centering{Verifiable Privacy-Preserving Aggregation Protocol}
            \begin{itemize}
                \item Step 0 (\textbf{Key Generation and Broadcast}):
			\vspace{0.5em}
   
                \textit{Client $C_u$: }
                \begin{itemize}
                    \item Generate public-private key pairs $(PK_u, SK_u)\leftarrow KeyGen(1^\lambda)$.
                    \item Send $PK_u$ to the server.
                \end{itemize}
                
                \textit{Server $S$: }
                \begin{itemize}
                    \item Collect more than $t$ messages from clients, who form a set \textbf{Client}$_1$. 
                    \item Assign each client in \textbf{Client}$_1$ with a unique index $u \in [1, |\textbf{Client}_1|]$. 
                    \item Broadcast $\{(u,PK_u)\}_{C_u\in\textbf{Client}_1}$ to all clients $C_u \in \textbf{Client}_1$.
                \end{itemize}
                
			\vspace{0.5em}
                \item Step 1 (\textbf{Shares Generation}):
			\vspace{0.5em}

                \textit{Client $C_u$: }
                \begin{itemize}
                    \item Receive the list $\{(u,PK_u)\}_{C_u\in\textbf{Client}_1}$ from the server.
                    \item Generate a $t$-$|\textbf{Client}_1|$ secret share $\{(s_{u,v}, o_{u,v})\}_{v\in \textbf{Client}_1}\leftarrow VSSGen(g,h, x_u, r_u, t, |\textbf{Client}_1|)$ for parameter $x_u$ with the Pedersen commitment $g^{x_u}h^{r_u}$, along with a set of commitments $\{E_i\}_{i\in[0,t-1]}$.
                    \item Encrypt the shares with corresponding public key $ci_{u,v}\leftarrow Enc((s_{u,v},o_{u,v}),PK_v)$ to obtain the ciphertext list $\{(v, ci_{u,v})\}_{v\in\textbf{Client}_1}$. 
                    \item Send the commitment list $\{E_{u,j}\}_{j\in[0,t-1]}$ and ciphertext list with receiver index $\{(v, ci_{u,v})\}_{v\in\textbf{Client}_1}$ to the server.
                \end{itemize}

                \textit{Server $S$: }
                \begin{itemize}
                    \item Collect more than $t$ messages from clients, who form a set $\textbf{Client}_2\subset \textbf{Client}_1$. 
                    \item Send all ciphertext with sender index $\{(u, ci_{u,v})\}_{v\in\textbf{Client}_2}$ to client $C_v$. 
                    \item Broadcast the benign client list $U_B\in\textbf{Client}_2$ obtained from the model filter to all clients $C_u\in\textbf{Client}_2$.
                \end{itemize}
                
			\vspace{0.5em}
                \item Step 2 (\textbf{Shares Aggregation and Reconstruction}):
			\vspace{0.5em}

                \textit{Client $C_u$: }
                \begin{itemize}
                    \item Receive the list $\{(v, ci_{v,u})\}_{v\in\textbf{Client}_2}$ and benign client list $U_B$ from the server.
                    \item Decrypt the shares $(s_{v,u}, o_{v,u})\leftarrow Dec(ci_{v,u}, SK_u)$ for $\{ci_{v,u}\}_{C_v\in U_B}$.
                    \item Aggregate the shares into global share $S_u = \Sigma_{C_v\in U_B} s_{v,u}$ and $O_u = \Sigma_{C_v\in U_B} o_{v,u}$.
                    \item Send the global share $(S_u, O_u)$ to the server. 
                \end{itemize}

                \textit{Server $S$: }
                \begin{itemize}
                    \item Collect more than $t$ messages from clients, who form a set $\textbf{Client}_3\subset\textbf{Client}_2$.
                    \item Reconstruct the global parameter $(X, R)\leftarrow VSSRec(t, \{S_u, O_u\}_{u\in\textbf{Client}_3})$ where $X=\Sigma_{C_u\in U_B}x_u$ and $R=\Sigma_{C_u\in U_B}r_u$.
                    \item Verify whether $\Pi_{C_u\in U_B} g^{x_u}h^{r_u} = g^Xh^R$. If the verification passes, continue the next round training, else, go to step x.
                \end{itemize}
                
			\vspace{0.5em}
                \item Step x (\textbf{Malicious Attacker Identification}):
			\vspace{0.5em}

                \textit{Server $S$: }
                \begin{itemize}
                    \item Verify each $\{(S_u, O_u)\}_{C_u\in \textbf{Client}_3}$ with $g^{S_u}h^{O_u} = \Pi_{j=0}^{t-1} {E_j^*}^{(u^j)}$, where $E_j^*=\Pi_{C_u\in U_B} E_{u,j}$. 
                    \item Send the commitment list with index $\{u, \{E_{u,j}\}_{j\in[0,t-1]}\}_{C_u \in U_B}$ to $C_u$ if the verification for $C_u$ fails.
                \end{itemize}

                \textit{Client $C_u$: }
                \begin{itemize}
                    \item Receive the commitment list $\{v, \{E_{v,j}\}_{j\in[0,t-1]}\}_{C_v \in U_B}$.
                    \item Verify each $\{(s_{v,u}, o_{v,u})\}_{C_v\in U_B}$ with $g^{s_{v,u}}h^{o_{v,u}}=\Pi_{j=0}^{t-1} {E_{v,j}}^{(u^j)}$. 
                    \item Generate a proof $\pi\leftarrow Prv\_Dec(PK_u, SK_u, (s_{v,u}, o_{v,u}), ci_{v,u})$ if the verification for $C_v$ fails.
                    \item Send the proof with index $(v, \pi)$ to the server.
                \end{itemize}

                \textit{Server $S$: }
                \begin{itemize}
                    \item Receive the proof $(v, \pi)$ from $C_u$.
                    \item Verify the proof $\pi=(m,A,B,z)$ with $g^e·A = g^z \land c_1^{e}\cdot B=(c_2m^{-1})^z$ where $(c_1,c_2)\leftarrow ci_{v,u}$, to obtain $result_1$. If $result_1$ fails, remove $C_u$ from all client lists and rebroadcast $U_B$ to all clients $C_u\in\textbf{Client}_2$ to re-execute Step 2.
                    \item Verify the share $(s_{v,u}, o_{v,u})\leftarrow m$ with $g^{s_{v,u}}h^{o_{v,u}}=\Pi_{j=0}^{t-1} {E_{v,j}}^{(u^j)}$ to obtain $result_2$.
                    \item If $result_1$ passes while $result_2$ not, remove $C_v$ from all client lists and rebroadcast $U_B$ to all clients $C_u\in\textbf{Client}_2$ to re-execute Step 2.
                \end{itemize}
            \end{itemize}
	}
    }
    \caption{The description of the verifiable privacy-preserving aggregation protocol in NoV.}
    \label{pre_peks}
\end{figure*}

\section{Security Analysis\label{Sec}}

In this section, we analyze the security of NoV, separated into the security of the aggregation protocol and the privacy of the whole process.

\subsection{Security}

First we illustrate the security of the model aggregation protocol, i.e., whether a malicious participant can break the aggregation protocol. 
The performance of the model filter will be illustrated by experiments in section \ref{Exp}.

With respect to the interaction flow of the model aggregation protocol, there are several malicious behaviors: 

\textbf{Generate an incorrect secret share in step 1.} 
In step 1, the client needs to submit the Pedersen commitment of the parameters for the verification of secret sharing, where it is required that the commitment used for secret-sharing verification is the same as the one previously used for model checking. 
The Pedersen commitment is computational binding~\cite{pedersen1991non}, where a probabilistic polynomial-time adversary cannot find $x'$ that satisfies $g^{x}h^{r}=g^{x'}h^{r'}\land x\neq x'$ with all but negligible probability. 
Thus the malicious client cannot swipe the values of the parameters corresponding to the commitments.
In step 1, the client needs to generate secret shares of the parameters. 
The Pedersen VSS is computational correctness~\cite{pedersen1991non}, where a probabilistic polynomial-time adversary $C_u$ cannot find $S_1, S_2$ of size $t$ that satisfies  $\{g^{s_{u,v}}h^{o_{u,v}}=\Pi_{j=0}^{t-1} {E_j}^ {(v^j)}|(s_{u,v},o_{u,v})\in S_1 \cup S_2 \land S_1\neq S_2\}$ with all but negligible probability. 
Thus the malicious client cannot generate shares that correspond to different parameters without being detected.
In step x, the honest client will submit a proof to the server when it receives an incorrect secret share. 
The $\Sigma$-protocol is perfect completeness~\cite{bootle2015short}, so upon receiving an incorrect secret share from the malicious client, the proof submitted by the honest client will be accepted by the server with certainty. 
Thus the malicious client cannot send an incorrect encrypted share.

\textbf{Submit an incorrect global share in step 2.}
In step 2, the client needs to aggregate multiple shares into one global share and send it to the server. 
The Pedersen VSS is computational correctness~\cite{pedersen1991non}, where a probabilistic polynomial-time adversary $A$ cannot find $S_1^*, S_2^*$ of size $t$ that satisfies $\{g^{S_u}h^{O_u} = \Pi_{j=0}^{t-1} {E_j^*}^{(u^j)}|(S_u,O_u)\in S_1^* \cup S_2^* \land S_1^*\neq S_2^*\}$ with all but negligible probability.
Thus the malicious client cannot generate shares corresponding to different parameters, i.e., it cannot submit a share that can pass the verification and disrupt the aggregation.

\textbf{Submit an incorrect proof in step x.}
In step x, the client being asked need to submit a proof to the server. 
The $\Sigma$-protocol is perfect 2-special soundness~\cite{bootle2015short}, where a probabilistic polynomial-time adversary $A$ cannot find $m'$ that satisfies $\{g^e·A = g^z \land c_1^{e}\cdot B=(c_2m'^{-1})^z|c_1=h^r, c_2=m·g^r, m\neq m'\}$ with all but negligible probability. 
Thus a malicious client cannot submit a correct proof about an honest client to the server, i.e., it cannot maliciously report an honest client.


\subsection{Privacy}
The NIZK scheme Bulletproof used in the model checking phase is perfect SHVZK~\cite{bulletproof}, thus no information about the parameters is revealed during the model checking process. 
The Pedersen VSS scheme used in the model aggregation phase is perfect privacy~\cite{pedersen1991non}. 
For any $S \subset {1,. . . , n}$ of size at most $t - 1$ and any $view_S$ $Prob$[D has secret s | views]$ = Prob$[D has secret s] for all $s \in \mathcal{Z}_q$.
Thus a collusion of less than t malicious clients would not result in the disclosure of information about the parameter.
The $\Sigma$-protocol used in the model aggregation phase is perfect SHVZK~\cite{bootle2015short}, so the $\Sigma$-protocol does not leak information about the private key of clients. 
The encryption scheme ElGamal used in the model aggregation phase is IND-CPA~\cite{bootle2015short}, so the encrypted share does not leak information about the parameters.

It is worth mentioning that during the model aggregation phase, client $C_u$ can maliciously submit a $\Sigma$-protocol proof to the server, resulting in the share of one specific client $C_v$ being exposed to the server. 
However, since the number of malicious clients is less than $t$, the server can only obtain at most $t-1$ shares of $C_v$, and since the Pedersen VSS scheme is perfect privacy~\cite{pedersen1991non}, this does not lead to the leakage of the parameters of $C_v$.


\section{Performance Analysis\label{Exp}}

In the experimental section, we test the defense performance against poisoning attacks as well as the operational efficiency of scheme NoV.

\subsection{Experimental Setup}

\textbf{Setup.} The experiments run on an Ubuntu 18.04 instance (i7-9750H @2.60Hz, 16GiB RAM, 4 vCPU).
We implement NoV in Rust 1.68.0 and train machine learning models in Python with PyTorch~\cite{paszke2019pytorch}. 
We use the elliptic curve-25519 from dalek-cryptography library~\cite{curve25519} for cryptographic computations.
We use the OOD Federated Learning framework~\cite{wang2020attack} for simulating the poisoning attack in federated learning.
We use the Bulletproof library~\cite{bulletproof} for the range proofs.

\textbf{Models and Datasets.} The datasets employed in our study consist of the widely recognized EMNIST~\cite{cohen2017emnist} and CIFAR-10~\cite{cifar10}.
We adopt a 3-layer convolutional neural network (CNN) for EMNIST, LeNet5 and ResNet20 for CIFAR10. 
The datasets of honest participants are non-iid.
For the tail attack case, we adopt a similar strategy to Wang et al.~\cite{wang2020attack}. 
For EMNIST, we selected the number "7" with a horizontal bar from the Ardis dataset~\cite{kusetogullari2020ardis}, modified its label to "2". 
For CIFAR-10, we modified the label of Southwest Airlines plane sample to "truck".

\textbf{Attacks.} The attacks considered in the experiments include untargeted attacks, targeted attacks, tail attacks, and PGD attacks. 
Madry et al. state that PGD attacks are one of the strongest first order attack~\cite{madry2017towards}, therefore we use the defense performance against PGD as the main measure to evaluate the defense strategies. 

\subsection{Defense Against Poisoning Attacks}

We first illustrate the defense performance of NoV against data poisoning attacks except PGD. 
Then we compare the defense performance of NoV and other schemes against PGD attacks. 
Finally we analyze the defense performance of NoV against PGD attacks under different settings.

\begin{figure}
    \centering
    \subfigure[CNN-EMNIST]{
    \includegraphics[width=0.22\textwidth]{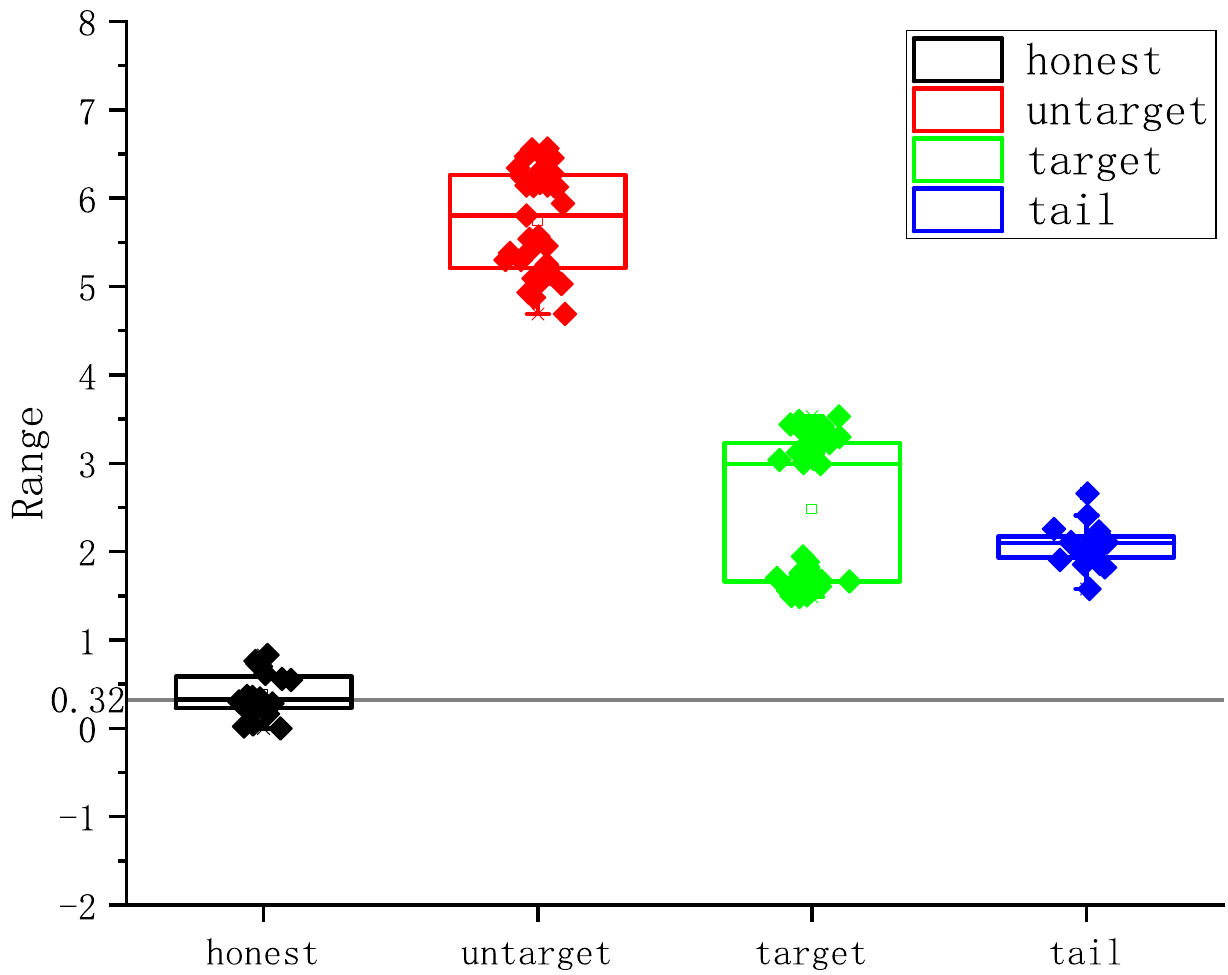}
    }
    \subfigure[LeNet5-CIFAR10]{
    \includegraphics[width=0.22\textwidth]{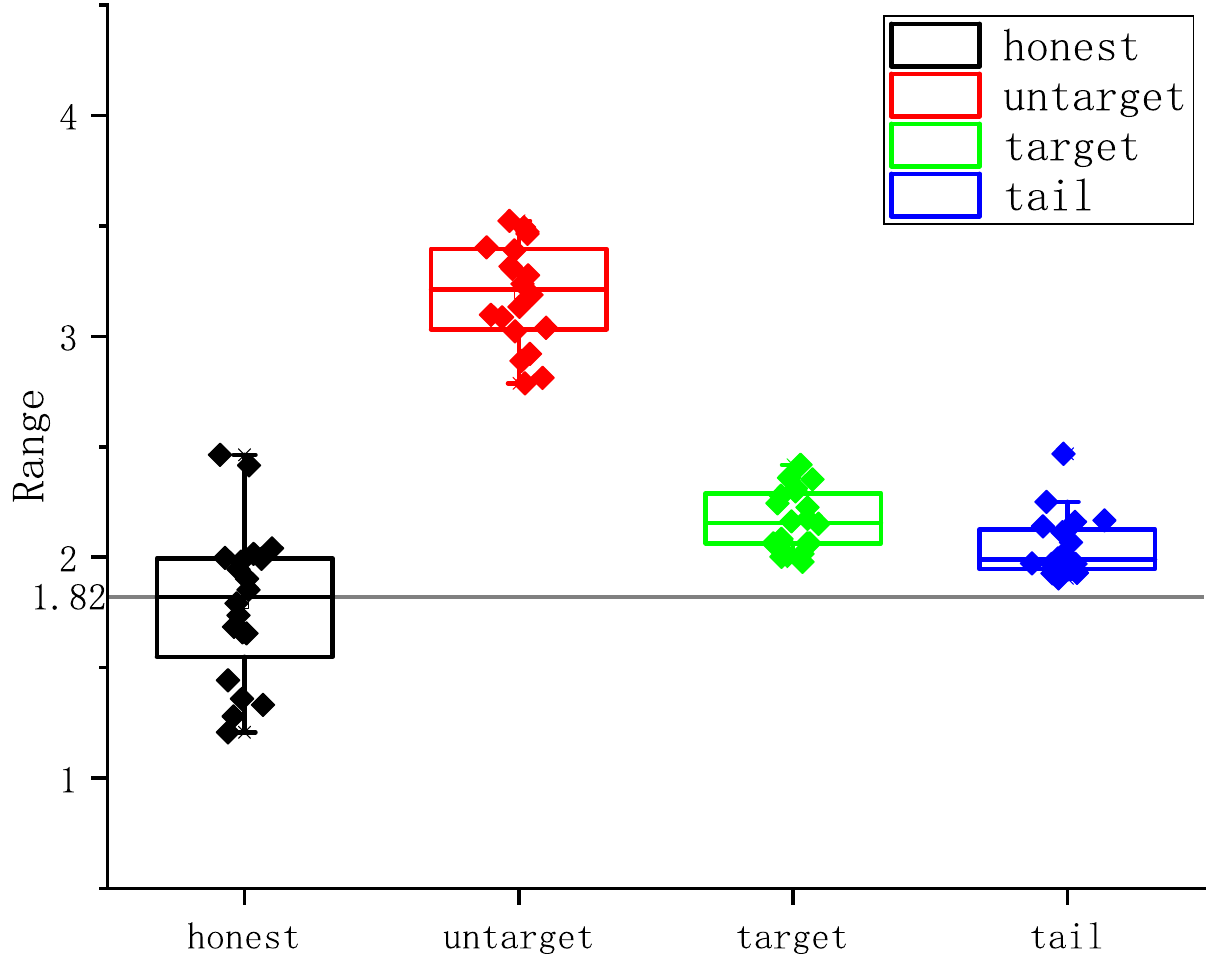}
    }
    \caption{Comparison of Euclidean distance among honest and malicious model updates on different models and tasks.}
    \label{dis_exp}
\end{figure}

Figure \ref{dis_exp} demonstrates the difference in magnitude between honest model updates and different kinds of malicious model updates. 
By setting an appropriate threshold $t_m$ (e.g., the median, 0.32 and 1.82, respectively), NoV can easily filter out these malicious model updates. 

\begin{figure*}
    \centering
    \subfigure[No Attack and No Defense]{
    \includegraphics[width=0.22\textwidth]{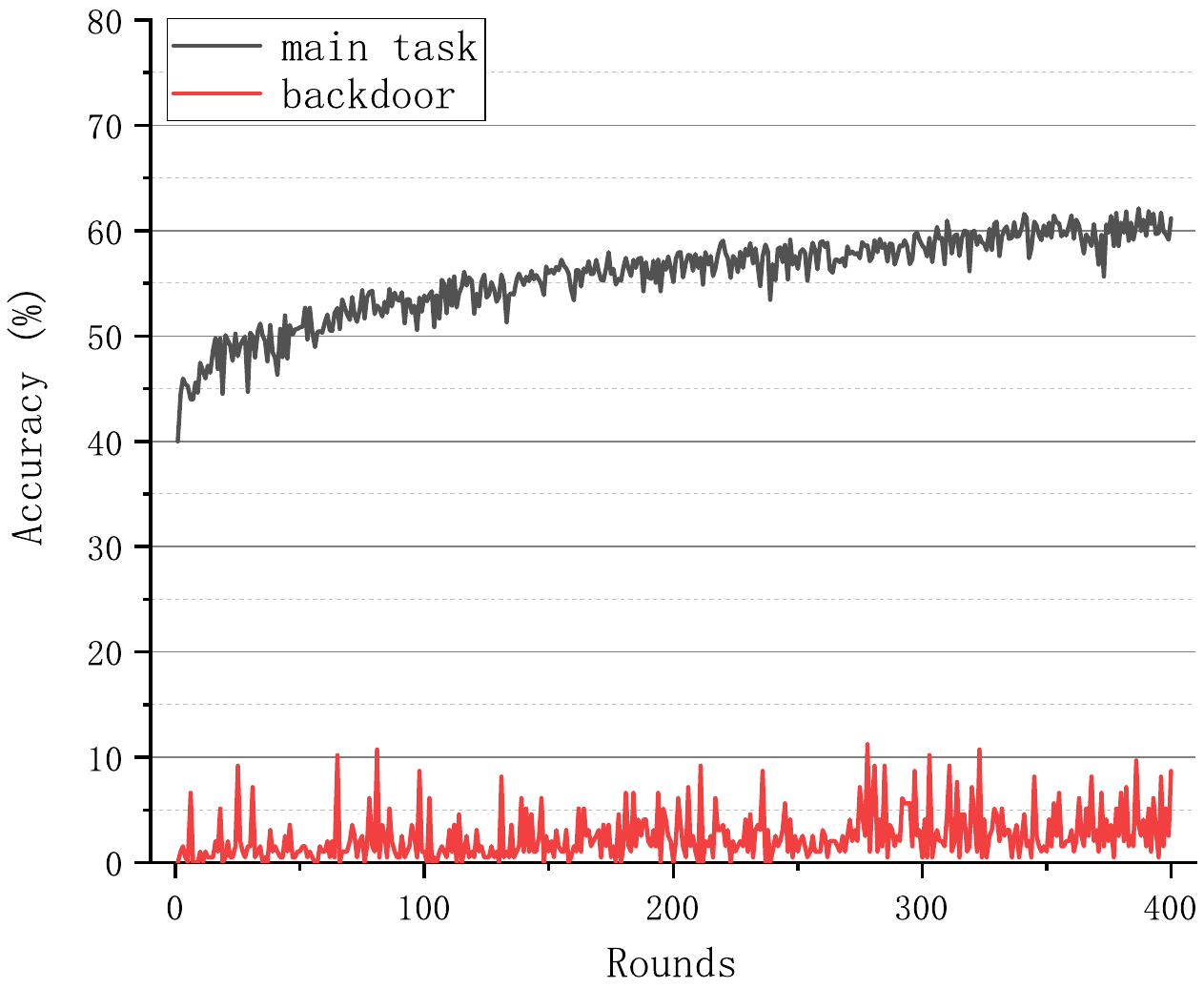}
    }
    \subfigure[PGD Attack and No Defense]{
    \includegraphics[width=0.22\textwidth]{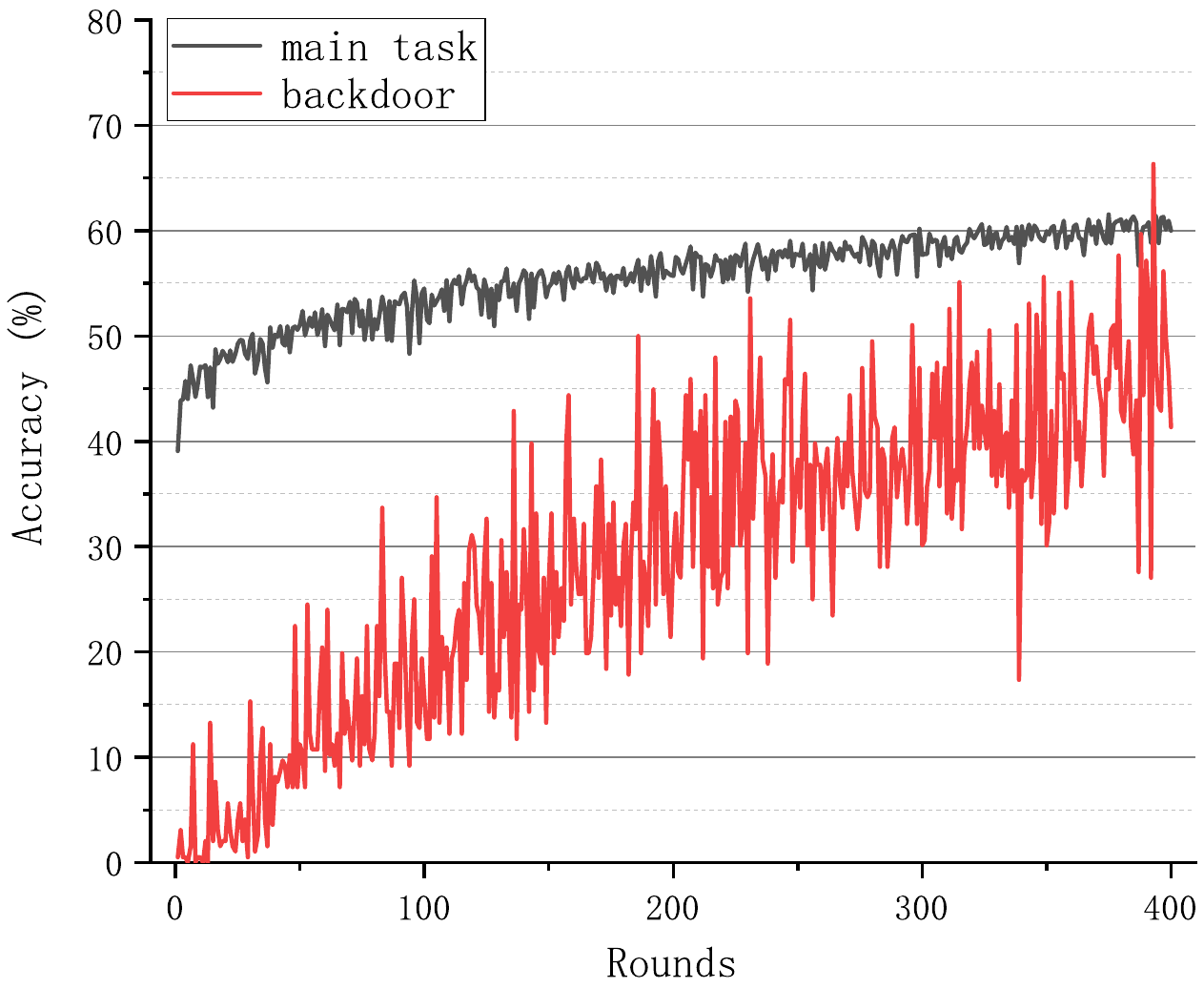}
    }
    \subfigure[NoV]{
    \includegraphics[width=0.22\textwidth]{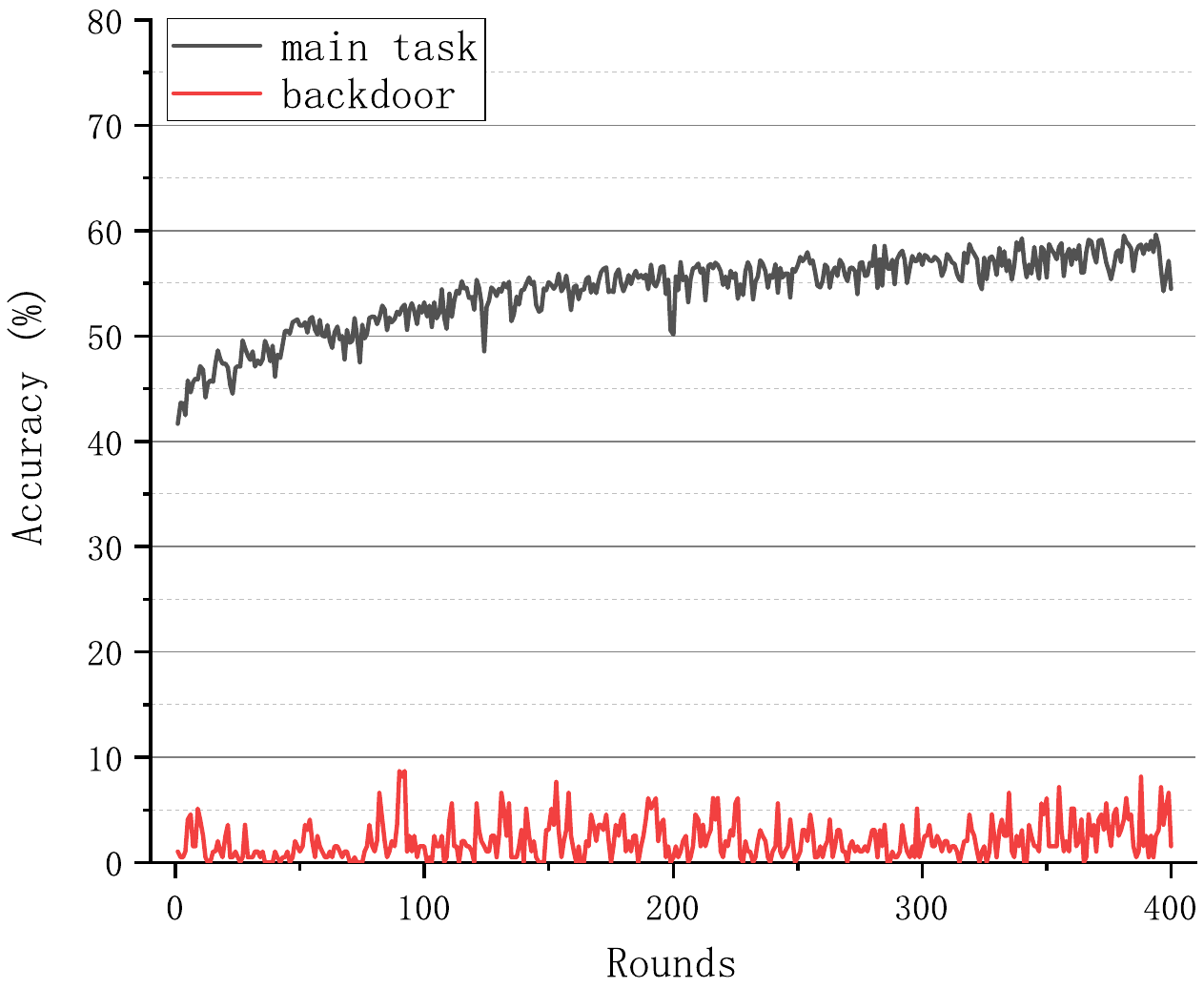}
    }\\
    \subfigure[FLTrust]{
    \includegraphics[width=0.22\textwidth]{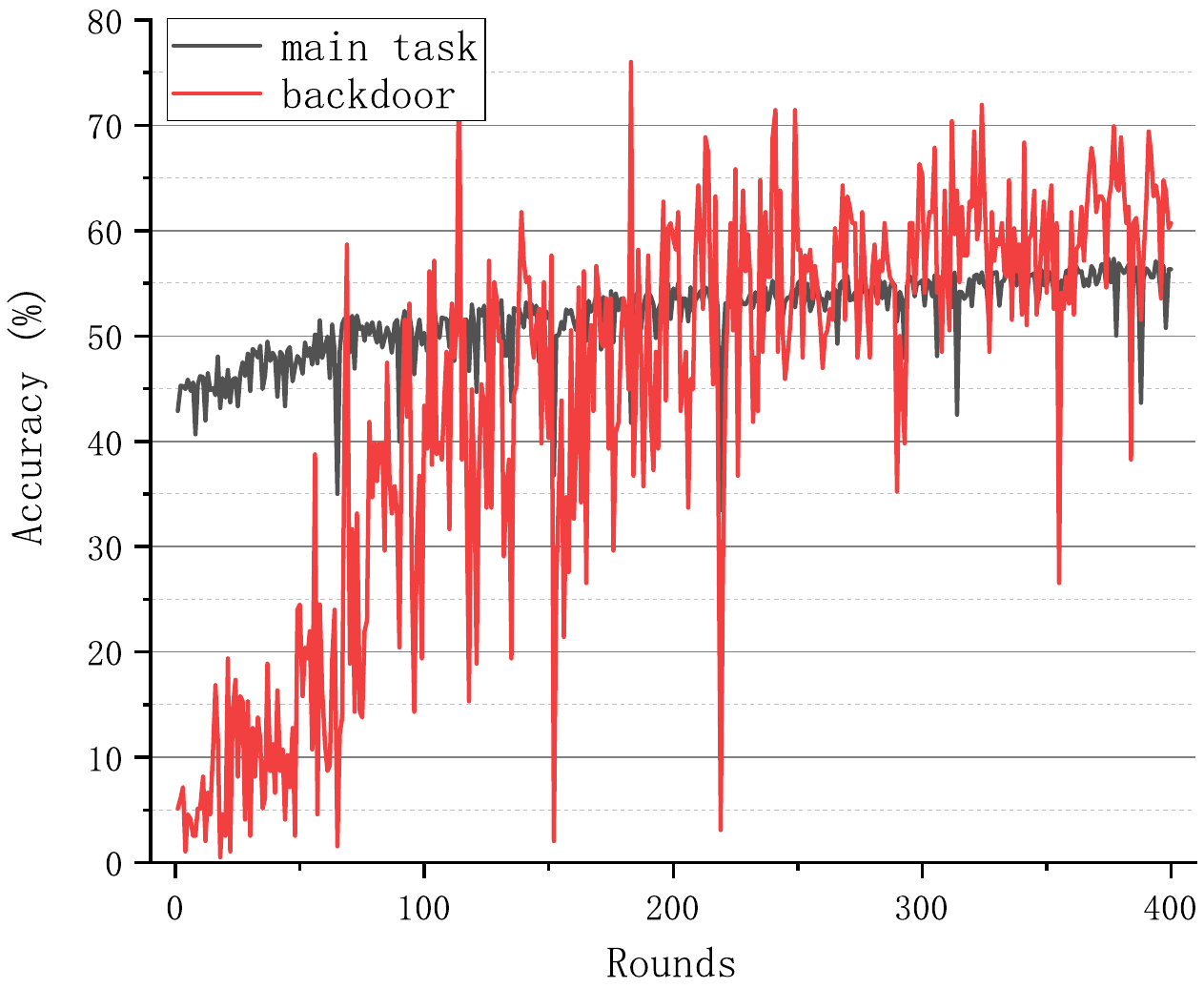}
    }
    \subfigure[CosDefense]{
    \includegraphics[width=0.22\textwidth]{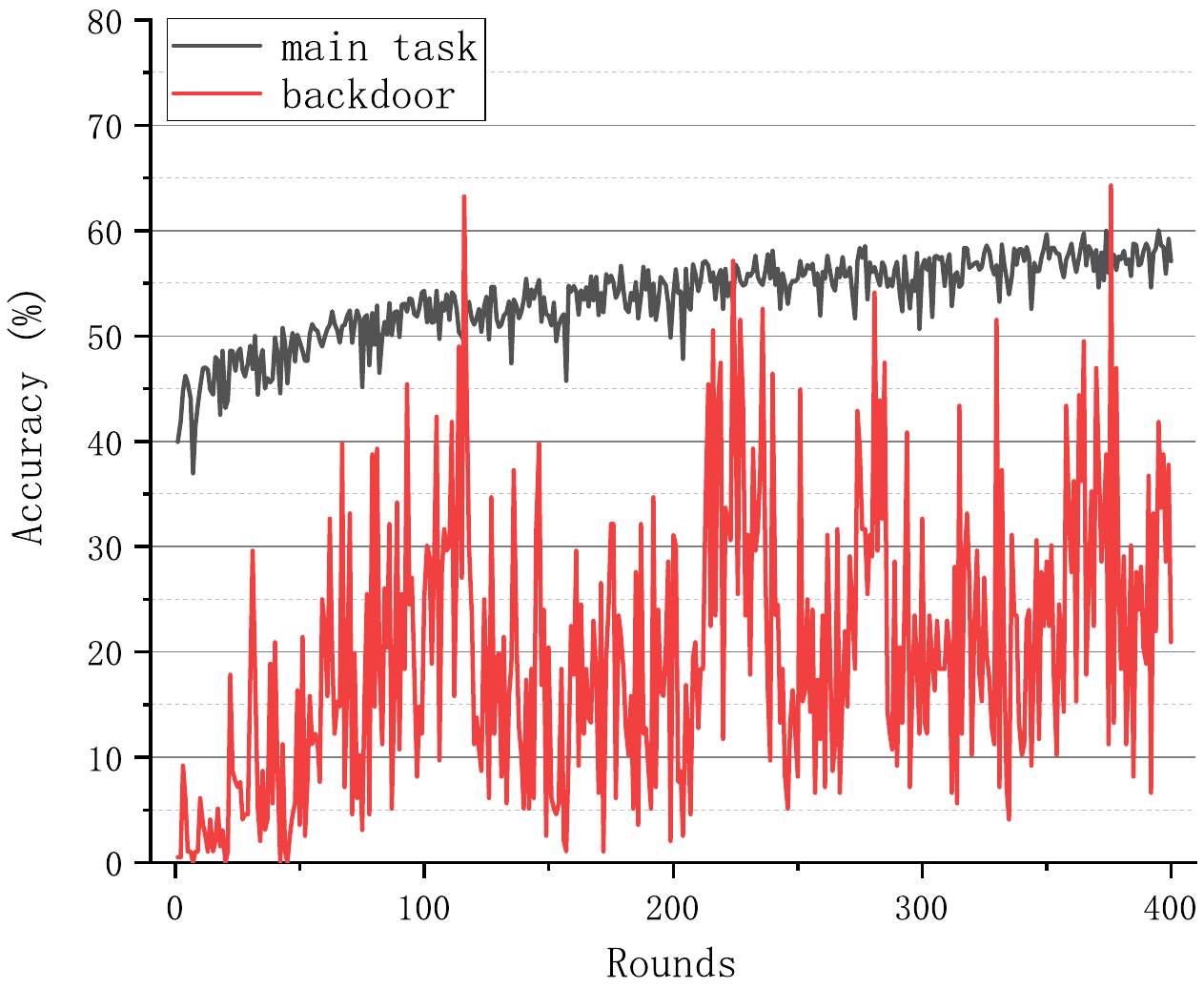}
    }
    \subfigure[RoFL $L_2$ norm]{
    \includegraphics[width=0.22\textwidth]{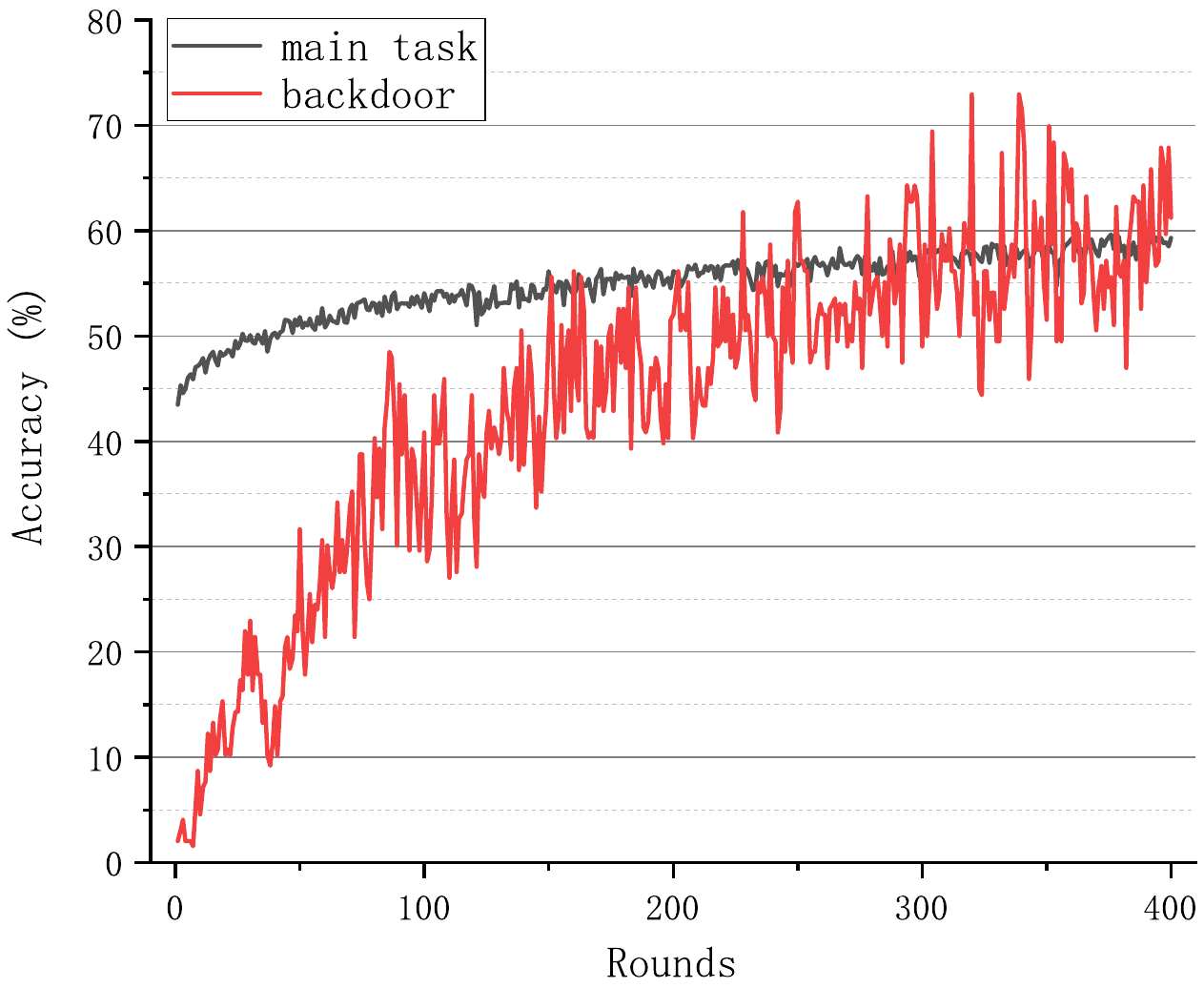}
    }
    \subfigure[RoFL $L_\infty$ norm]{
    \includegraphics[width=0.22\textwidth]{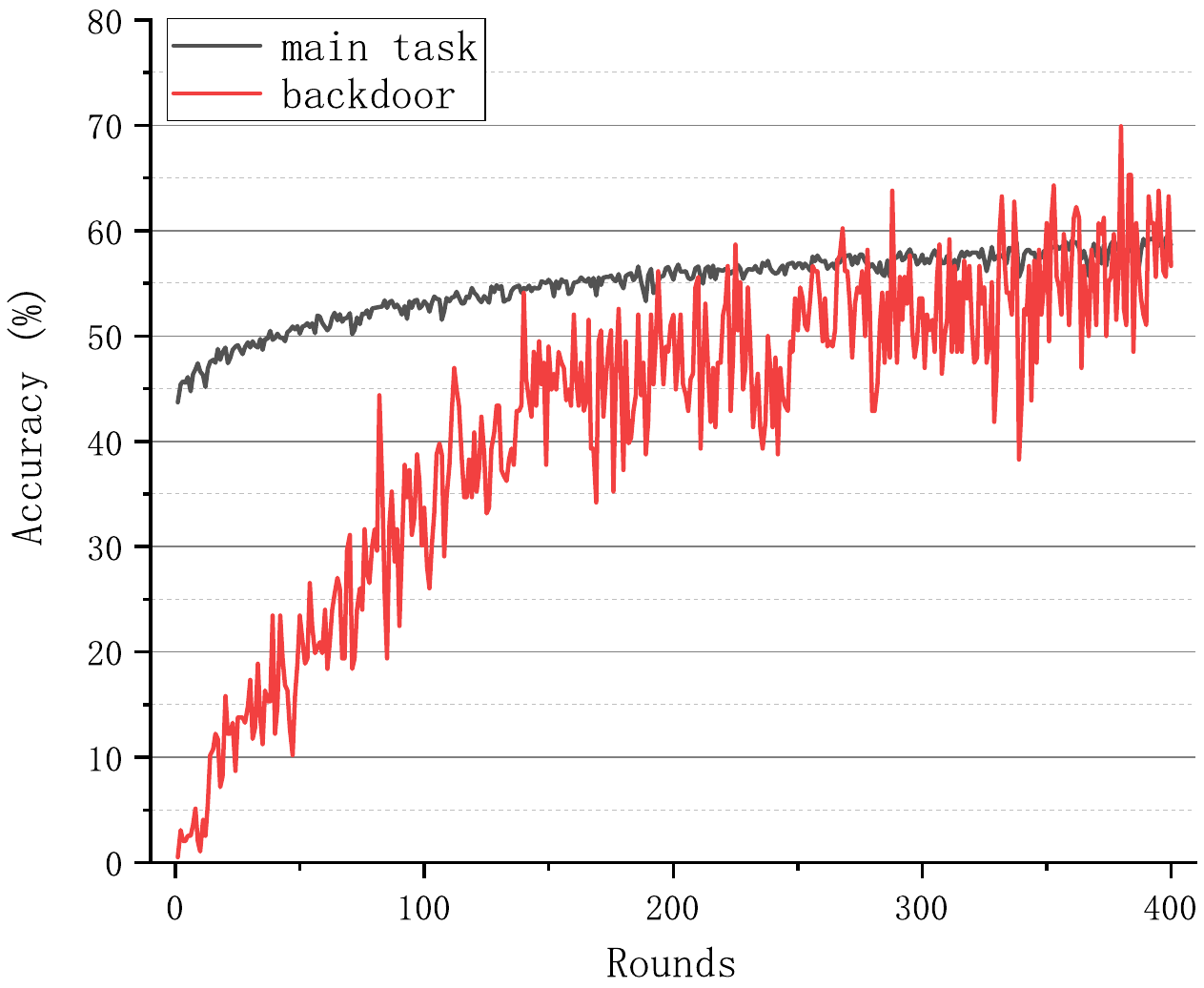}
    }
    \caption{Comparison among NoV and latest model filtering strategies defending against PGD attack on CIFAR10 with LeNet5. The black line represents the accuracy of the main task and the red line represents the accuracy of the backdoor task.}
    \label{comparecifar10_exp}
\end{figure*}

\begin{figure*}
    \centering
    \subfigure[No Attack]{
    \includegraphics[width=0.22\textwidth]{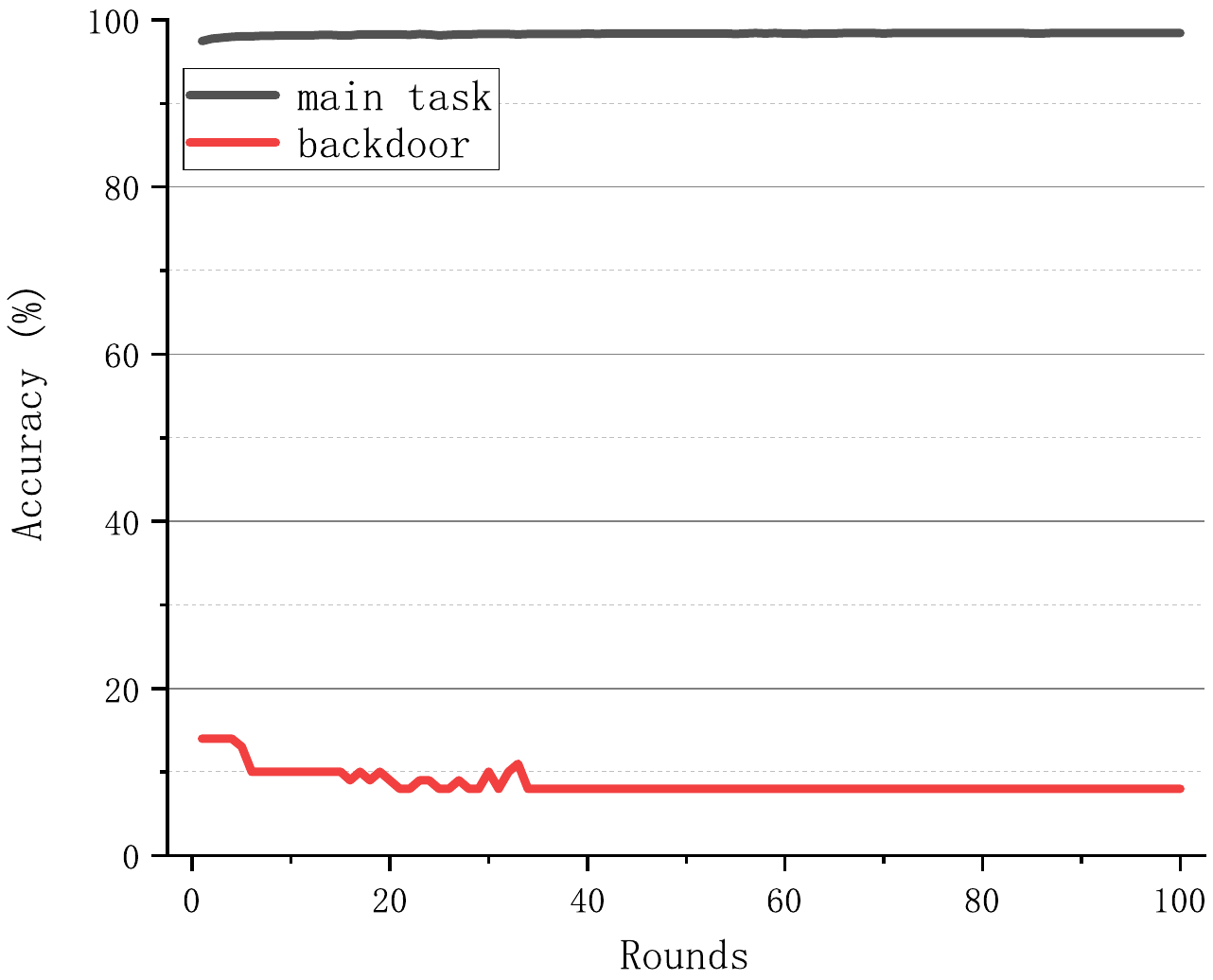}
    }
    \subfigure[No Defense]{
    \includegraphics[width=0.22\textwidth]{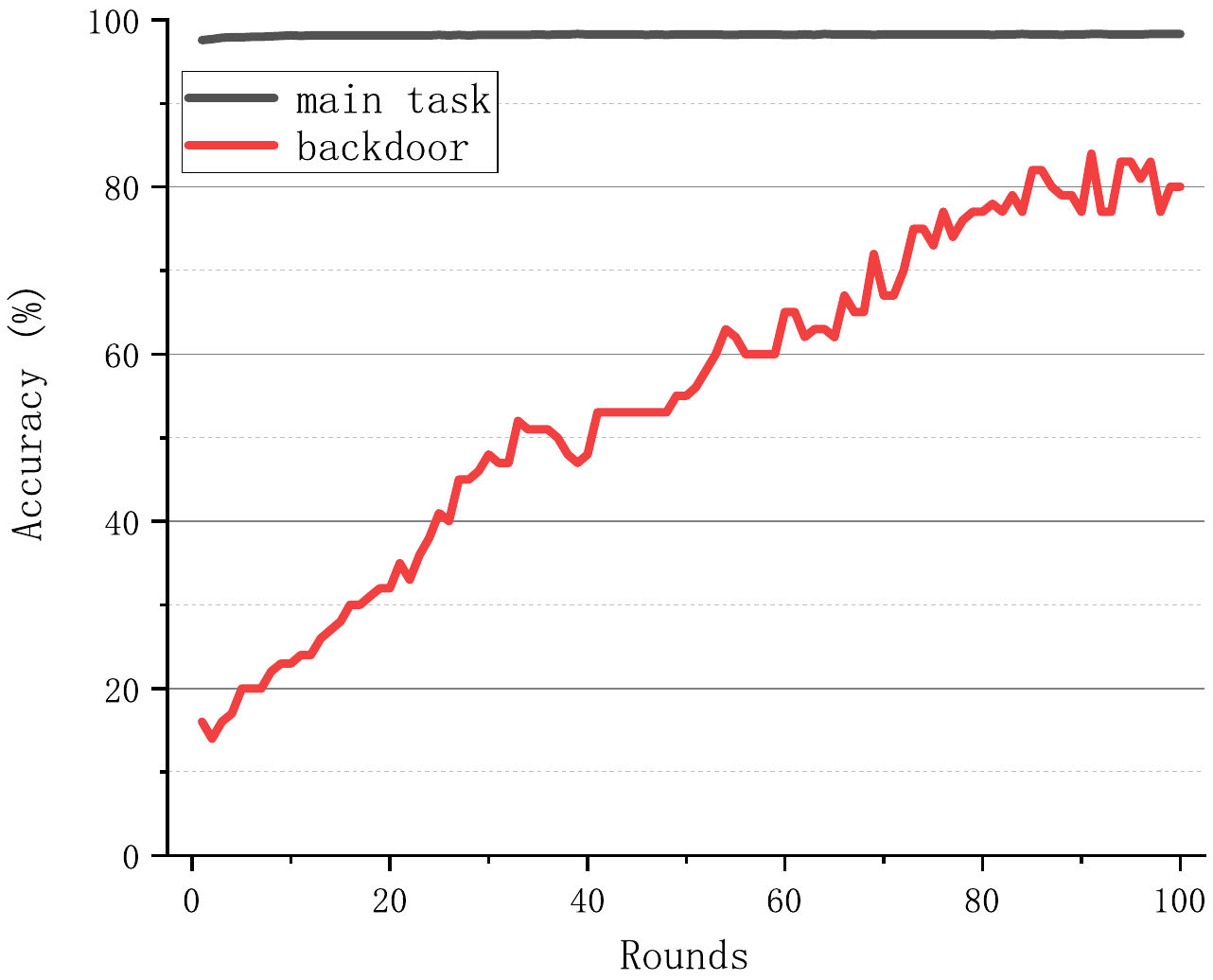}
    }
    \subfigure[NoV]{
    \includegraphics[width=0.22\textwidth]{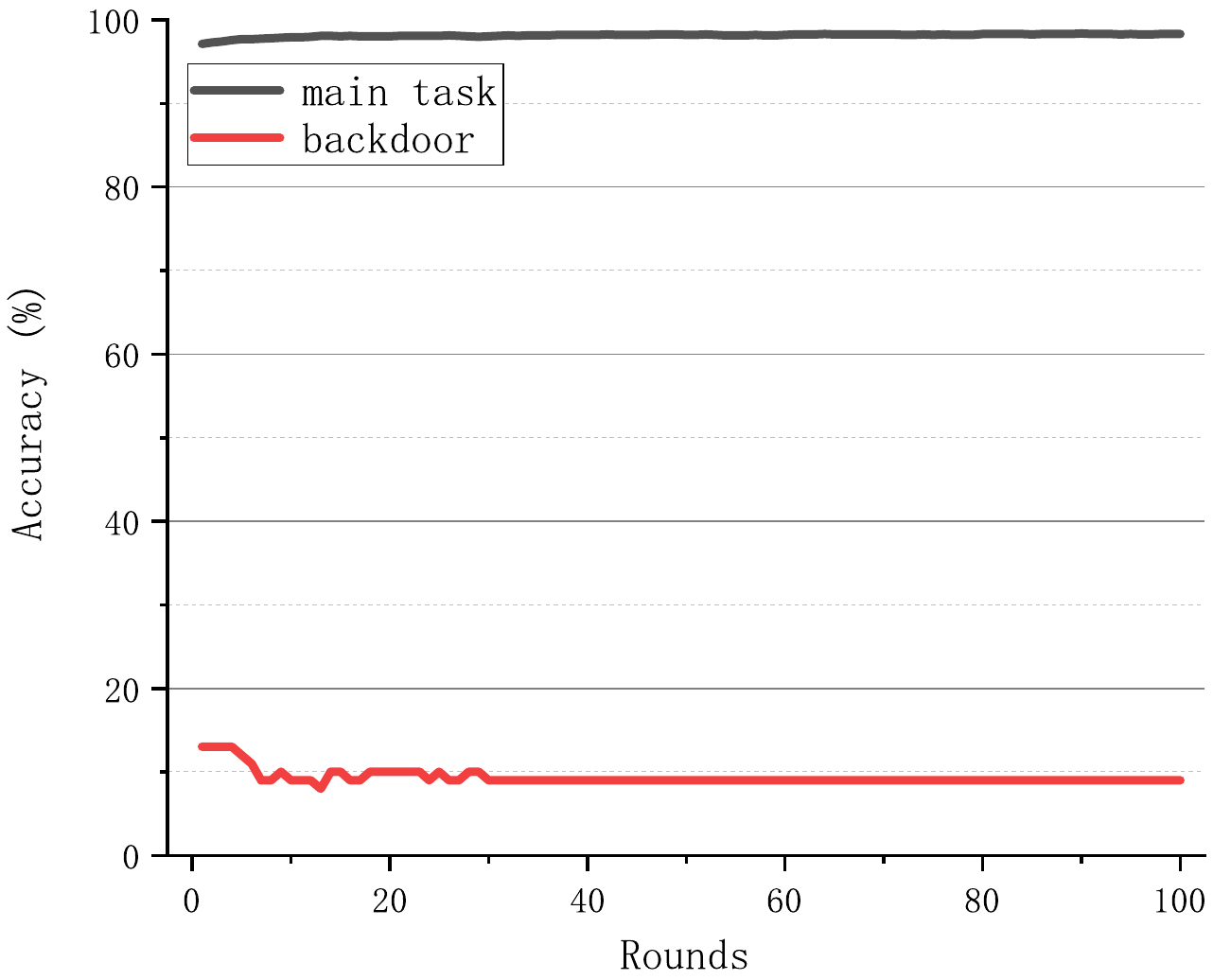}
    }\\
    \subfigure[FLTrust]{
    \includegraphics[width=0.22\textwidth]{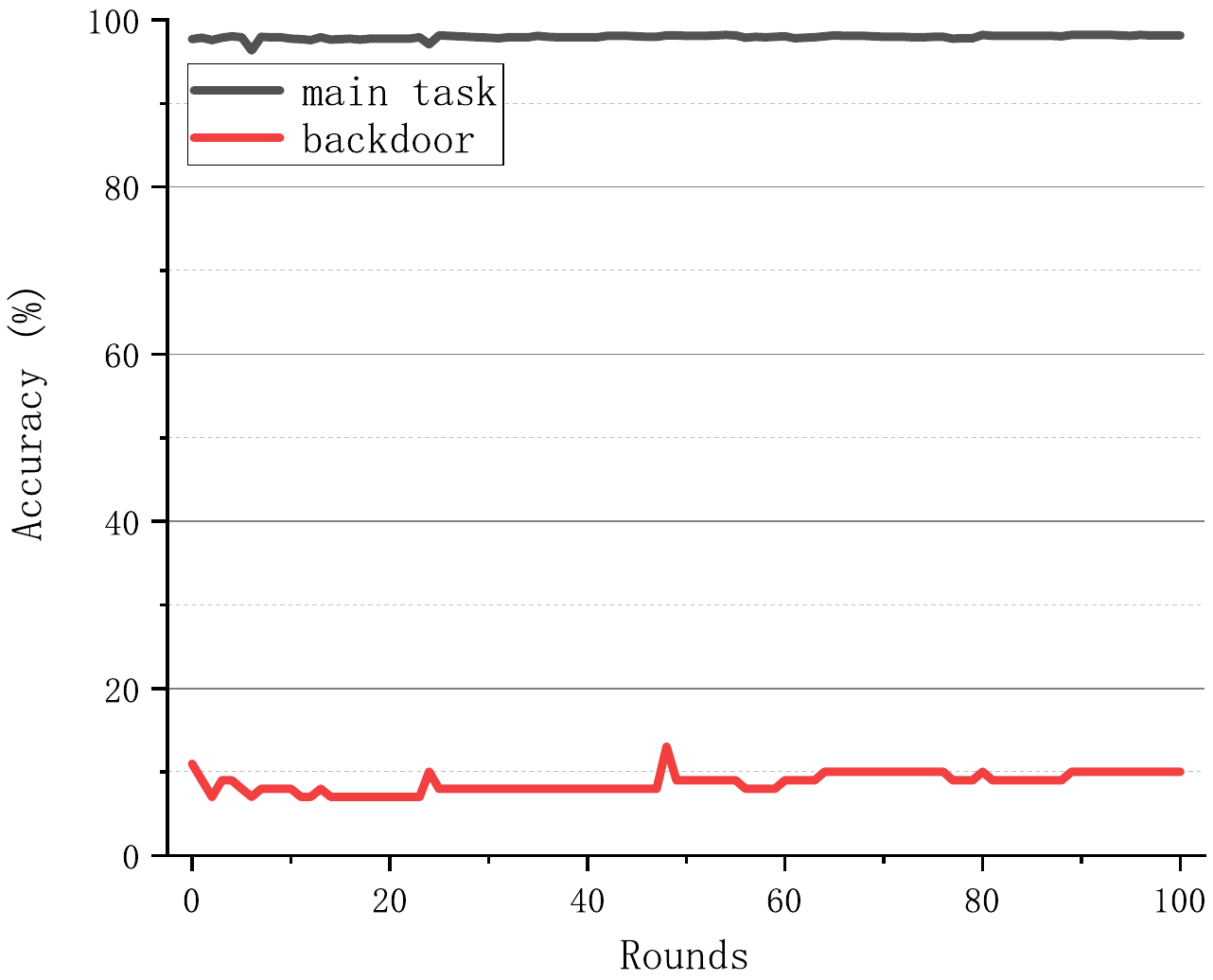}
    }
    \subfigure[CosDefense]{
    \includegraphics[width=0.22\textwidth]{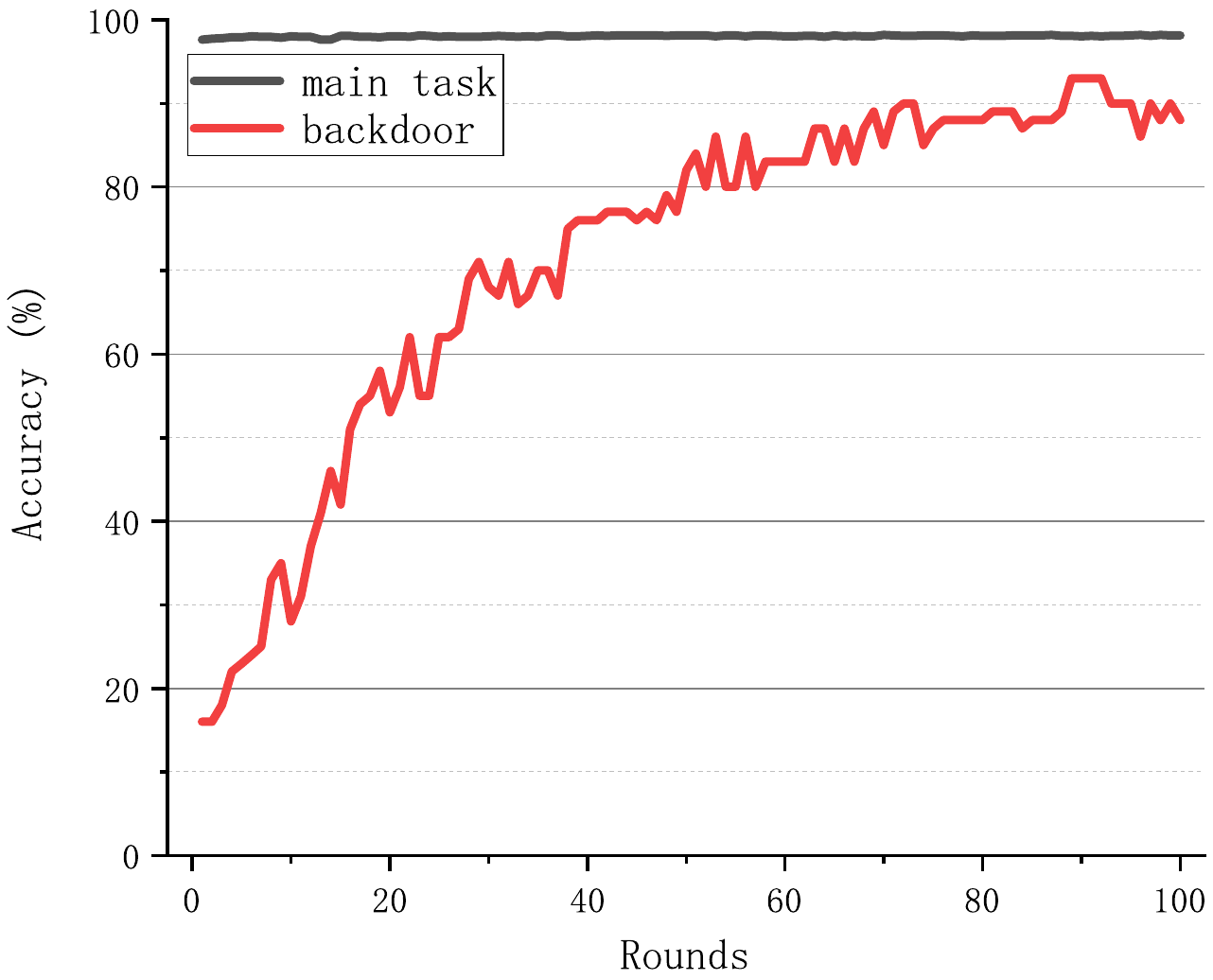}
    }
    \subfigure[RoFL $L_2$ norm]{
    \includegraphics[width=0.22\textwidth]{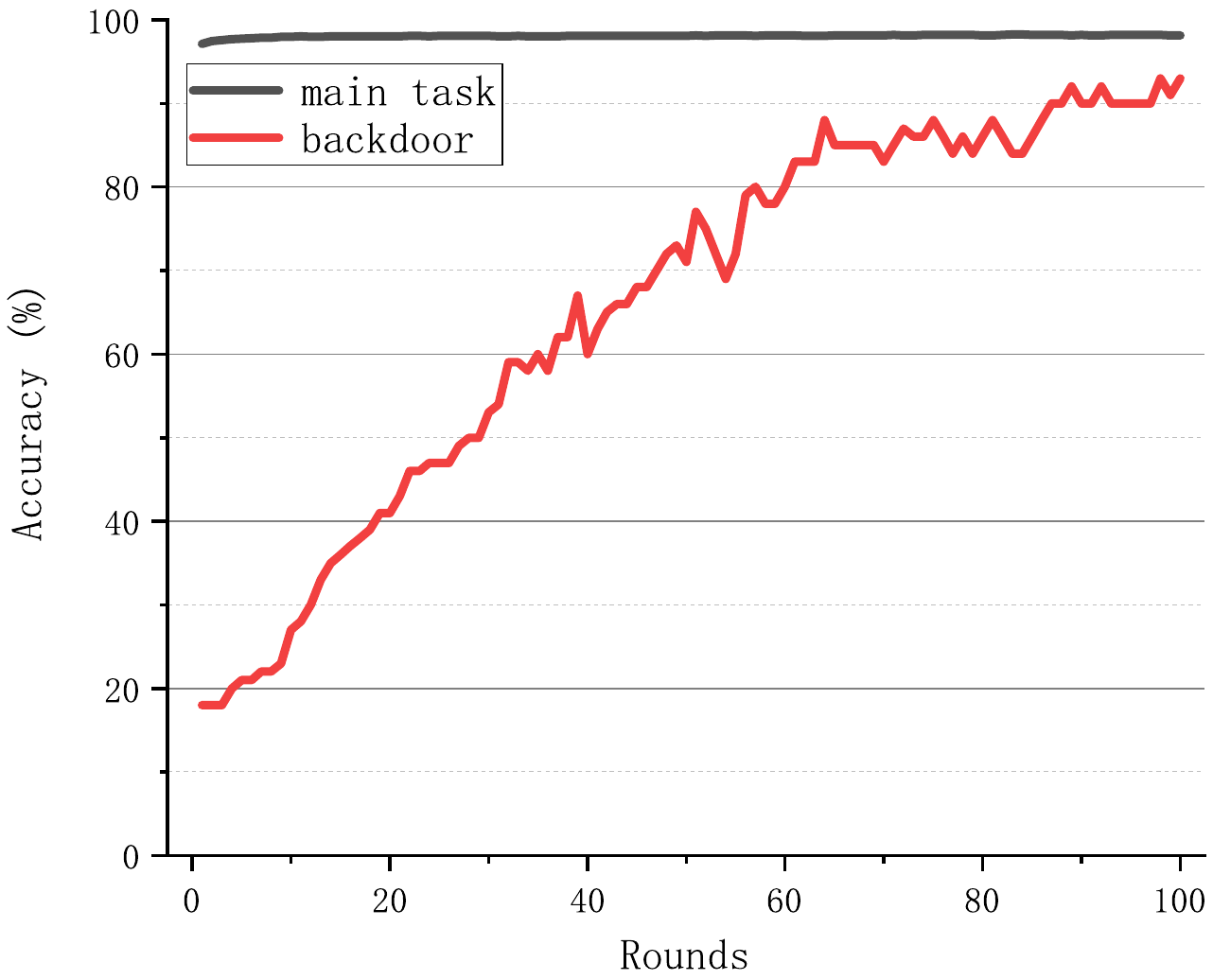}
    }
    \subfigure[RoFL $L_\infty$ norm]{
    \includegraphics[width=0.22\textwidth]{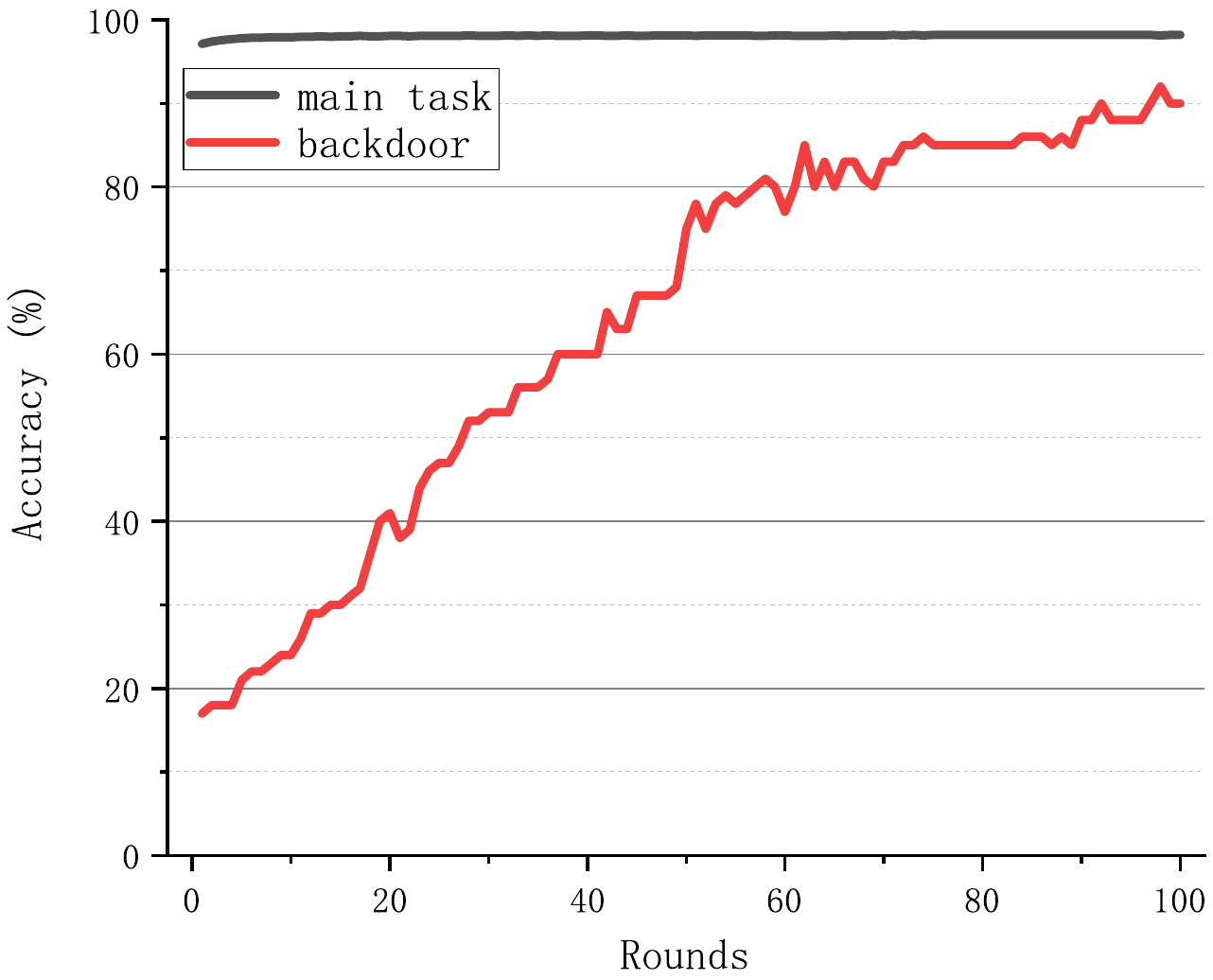}
    }
    \caption{Comparison among NoV and latest model filtering strategies defending against PGD attack on EMNIST with a 3-layer CNN. The black line represents the accuracy of the main task and the red line represents the accuracy of the backdoor task.}
    \label{compareemnist_exp}
\end{figure*}

With a public known threshold $t_m$, the PGD attack can bypass the first filtering strategy. 
Under this situation, we compare the defense performance of NoV with other schemes against PGD tail attack, for the poisoning with tail data is more effective~\cite{lycklama2023rofl}. 
We compare the performance of NoV with the latest model checking strategies (FLTrust, cosDefense, RoFL) under PGD tail attacks. 
These experiments are conducted with 30 trainers containing 3 attackers all with non-iid datasets.
For CIFAR10, the threshold $t_m$ and $t_s$ are set to 1.0 and 0.4 respectively. 
For EMNIST, the threshold $t_m$ and $t_s$ are set to 0.2 and 0.5 respectively. 
The results are shown in figure \ref{comparecifar10_exp} and \ref{compareemnist_exp}.
It can be seen that NoV is able to maintain a low backdoor task accuracy without affecting the main task accuracy with different tasks and datasets, which demonstrates that NoV can effectively defend against PGD attacks in the non-iid case. 

Regarding the defense performance of other strategies, FLTrust is able to defend against PGD on EMNIST.
However, FLTrust is not able to identify the attacker on CIFAR10, and instead filters out a portion of the honest participants, resulting in an increased backdoor accuracy compared to when there is no defense.
CosDefense is able to filter out PGD attackers on CIAFR10 to a some extent, decreasing the backdoor accuracy. 
But nearly half of the honest participants is also filtered out in this process. 
On EMNIST, CosDefense can only filter a few PGD attackers, while excluding nearly half of the honest participants, resulting in a higher backdoor accuracy than there is no defense.
Considering the defense strategy of RoFL and the publicly available threshold $t_m$, we project both the honest and malicious updates onto the given threshold. 
Thus RoFL with $L_2$ norm is unable to filter out any update for the public constraints. 
Besides, RoFL with $L_\infty$ also filters no malicious updates because they have smaller $L_\infty$.
Furthermore, due to the projection, the magnitude of the honest participant's update decreases, making the impact of poisoning relatively larger, which in turn improves the backdoor accuracy.

\begin{figure*}
    \centering
    \subfigure[1 Attacker (3\%)]{
    \includegraphics[width=0.22\textwidth]{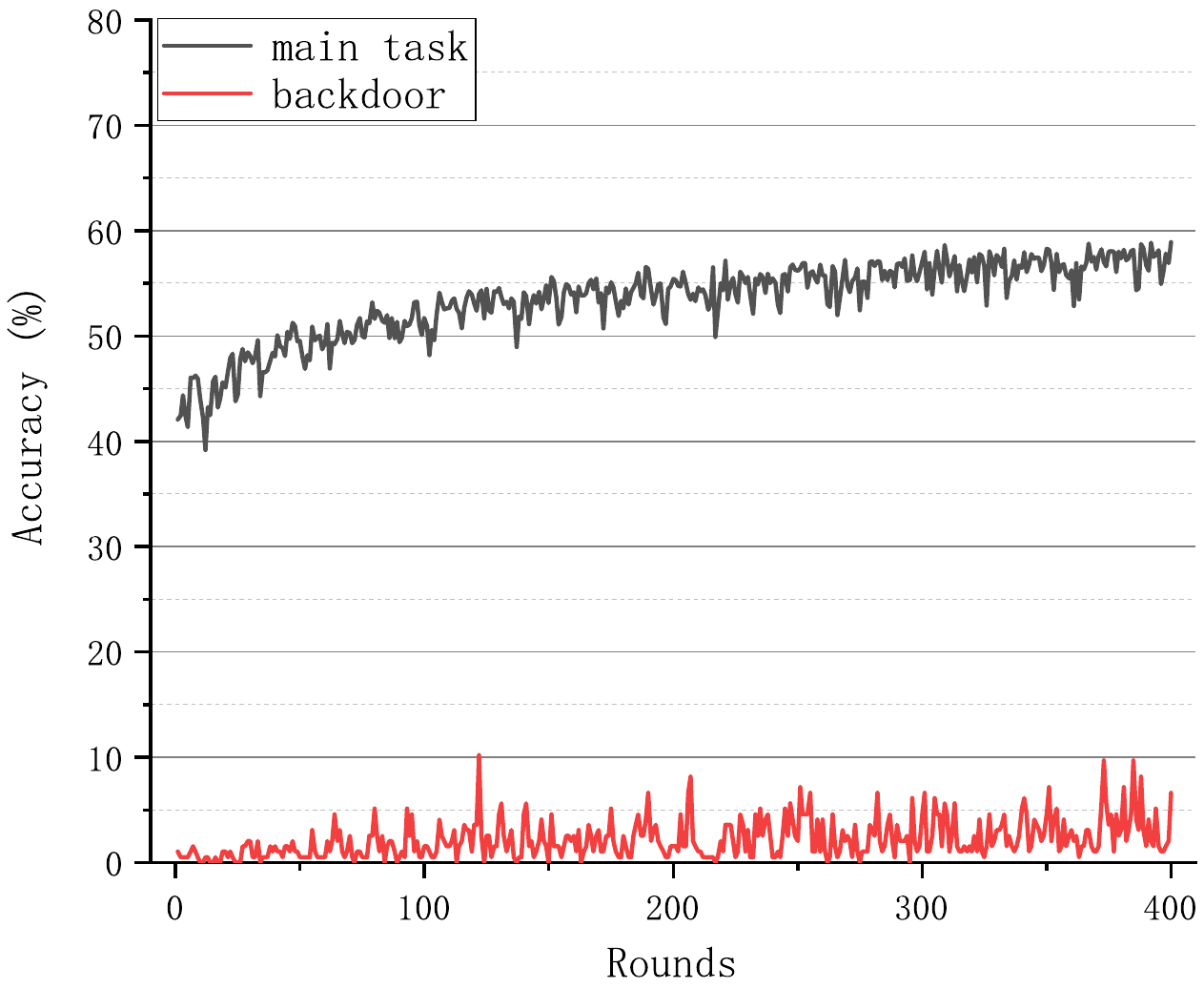}
    }
    \subfigure[3 Attackers (10\%)]{
    \includegraphics[width=0.22\textwidth]{expc/nov_cifar10.pdf}
    }
    \subfigure[5 Attackers (17\%)]{
    \includegraphics[width=0.22\textwidth]{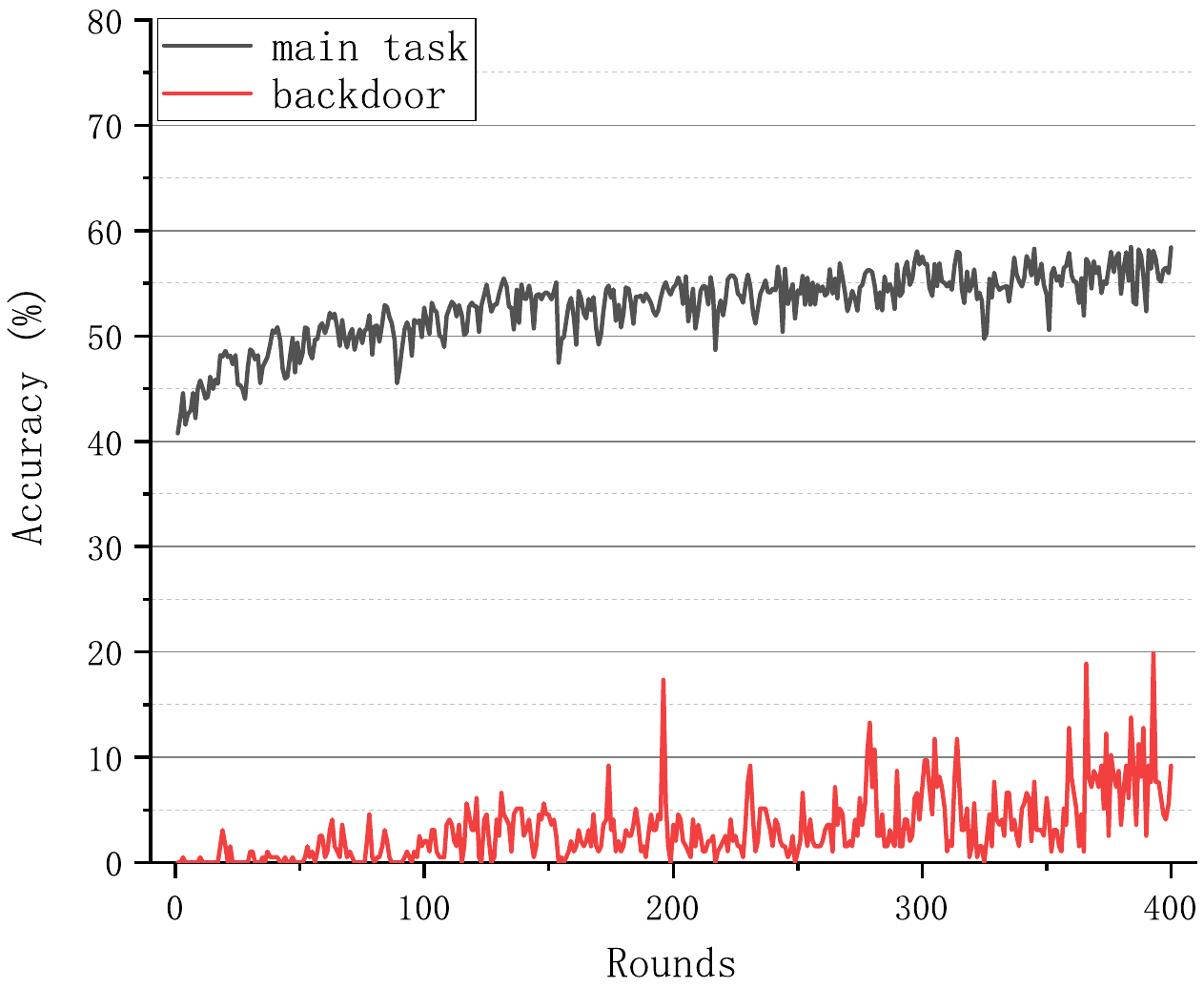}
    }
    \subfigure[10 Attackers (33\%)]{
    \includegraphics[width=0.22\textwidth]{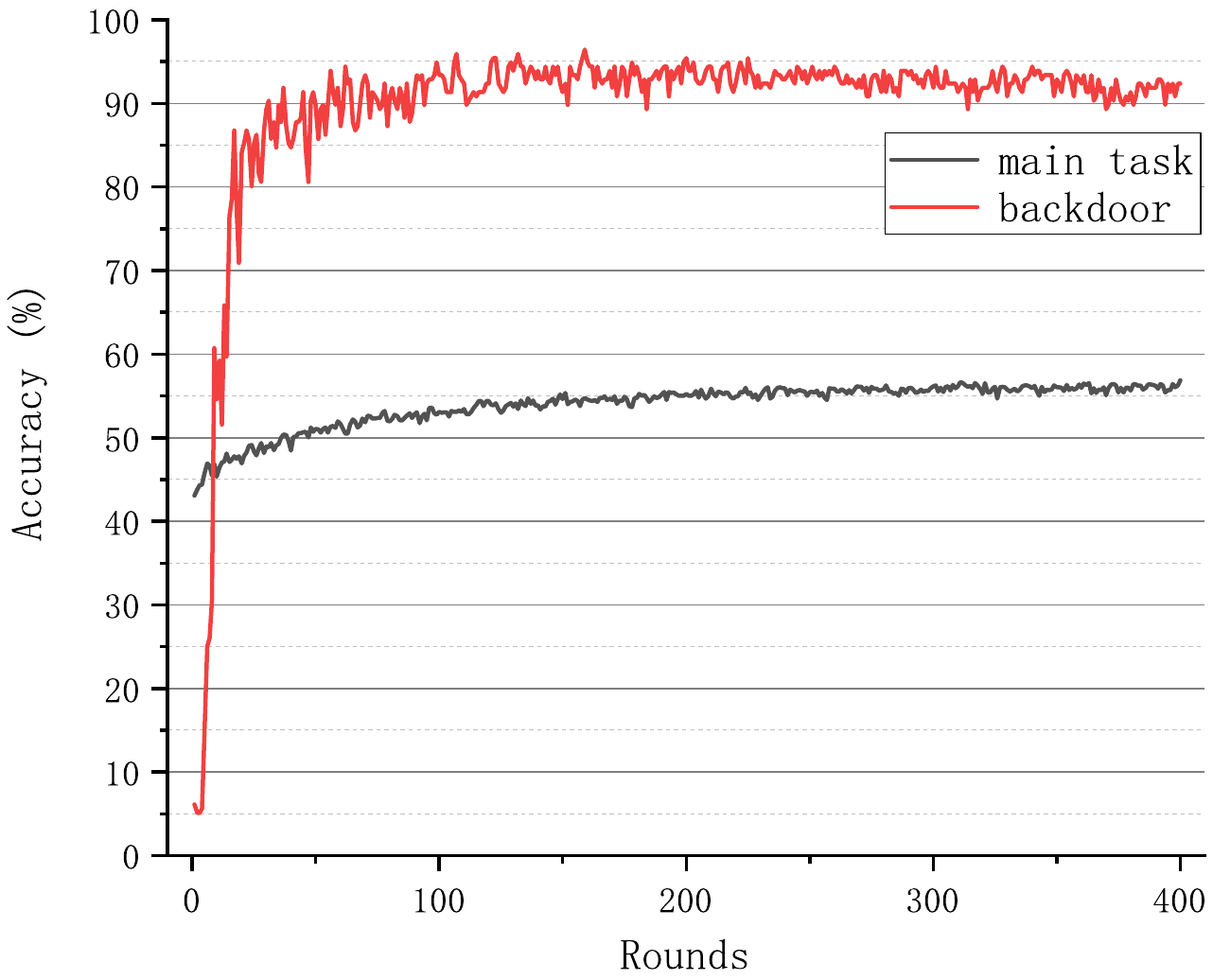}
    }
    \caption{Comparison of the defense performance of NoV with different number of attackers on CIFAR10 with LeNet5. The black line represents the accuracy of the main task and the red line represents the accuracy of the backdoor task.}
    \label{novcifar10_exp}
\end{figure*}

\begin{figure*}
    \centering
    \subfigure[1 Attacker (3\%)]{
    \includegraphics[width=0.22\textwidth]{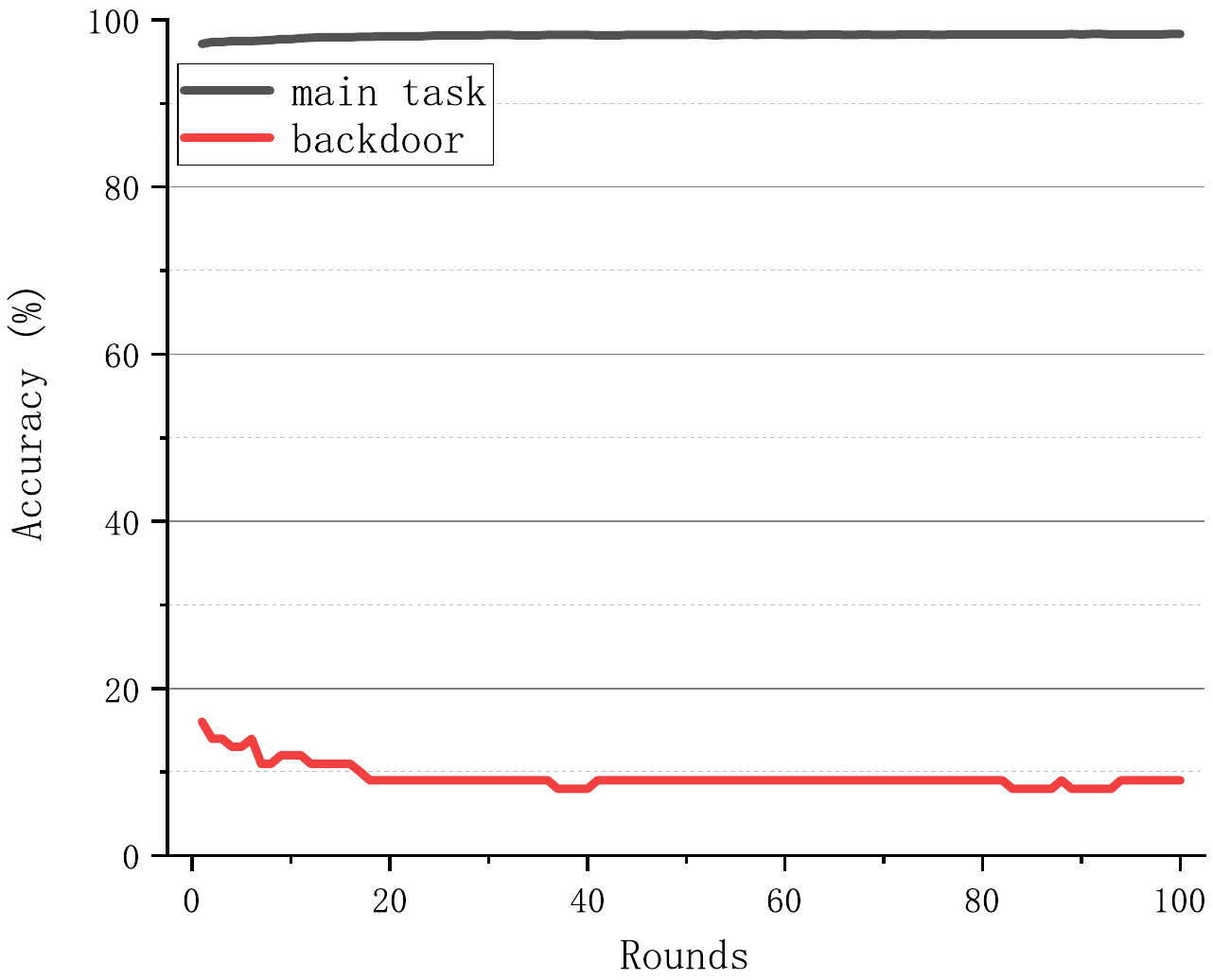}
    }
    \subfigure[3 Attackers (10\%)]{
    \includegraphics[width=0.22\textwidth]{expc/nov_emnist.pdf}
    }
    \subfigure[5 Attackers (17\%)]{
    \includegraphics[width=0.22\textwidth]{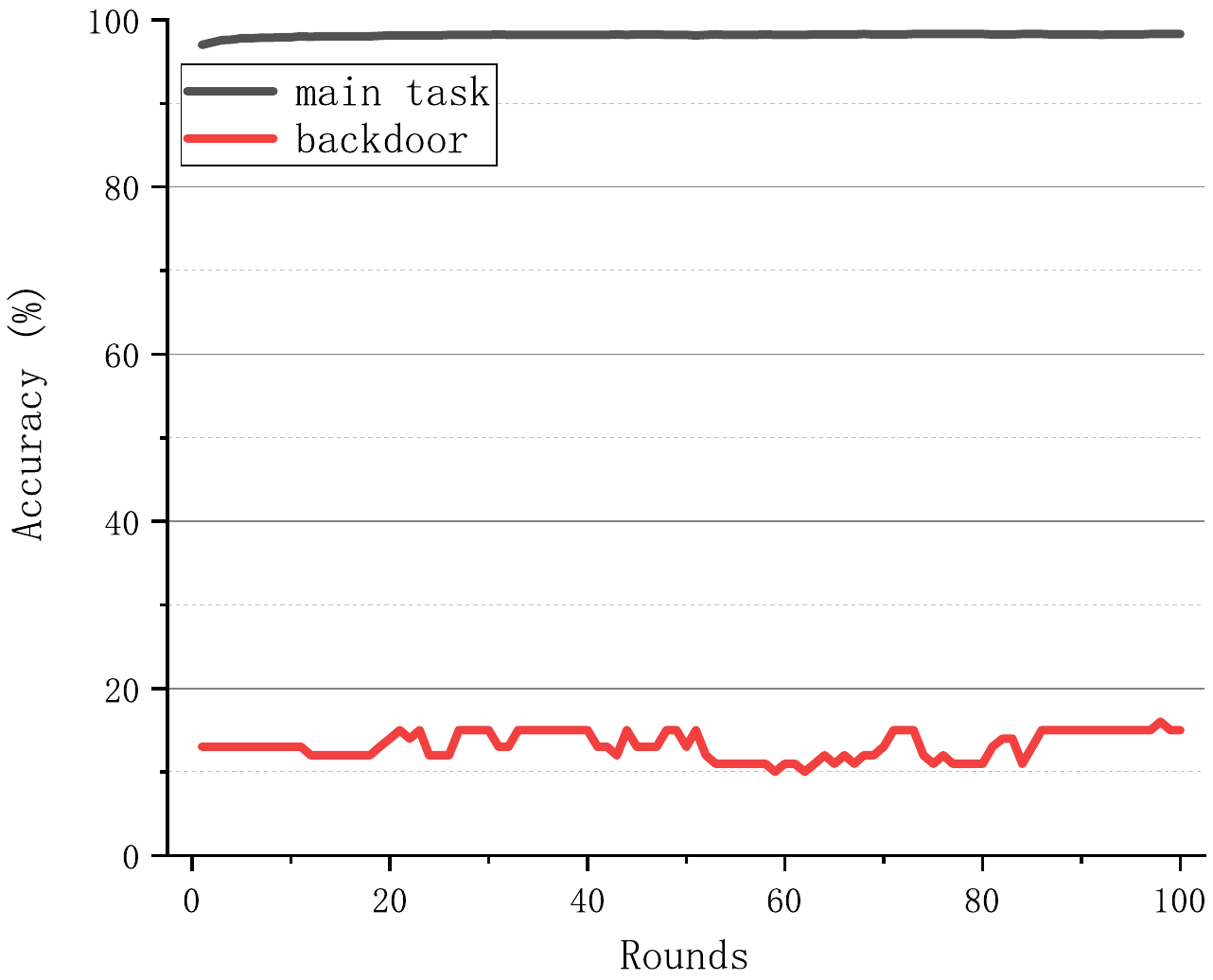}
    }
    \subfigure[10 Attackers (33\%)]{
    \includegraphics[width=0.22\textwidth]{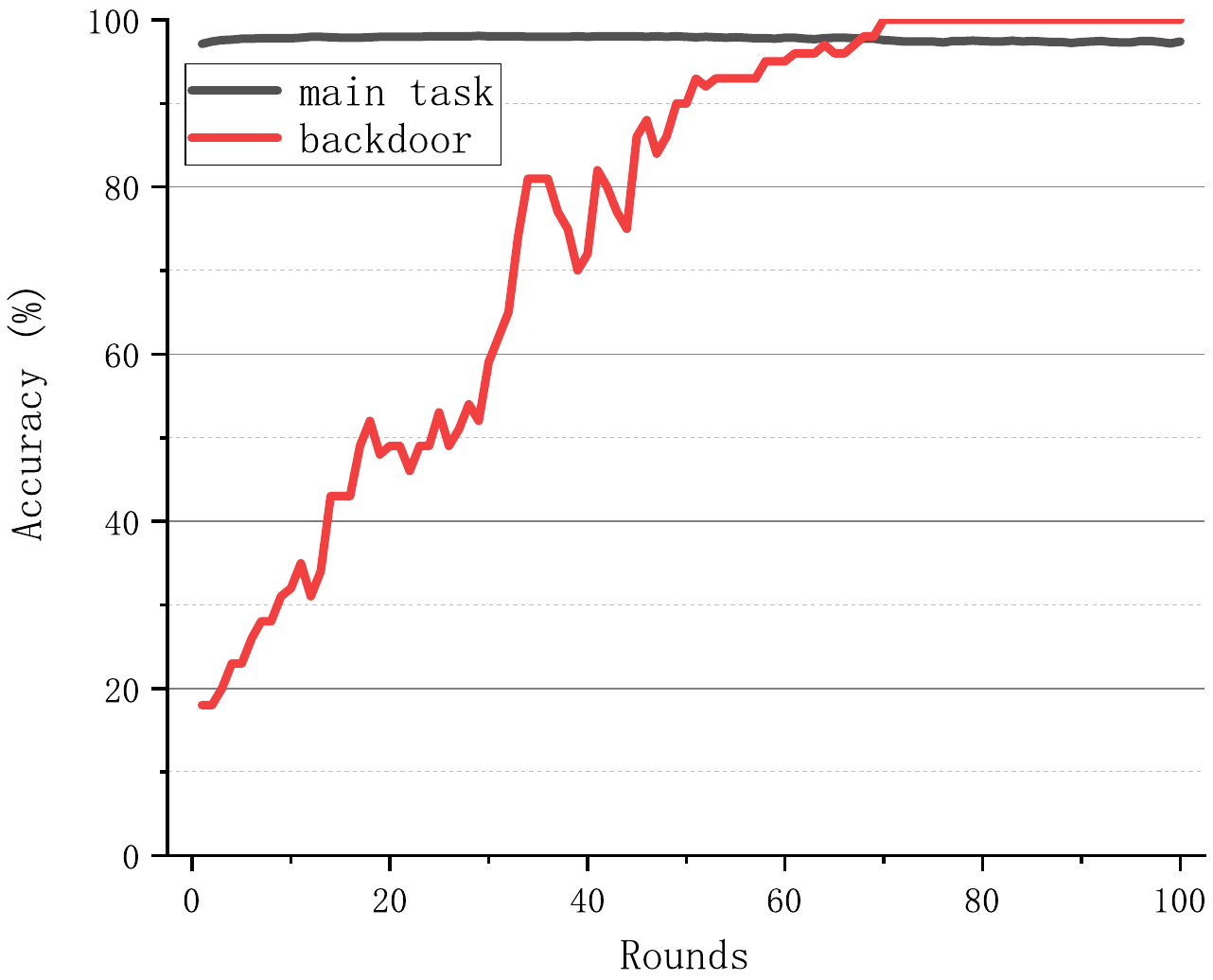}
    }
    \caption{Comparison of the defense performance of NoV with different number of attackers on EMNIST with a 3-layer CNN. The black line represents the accuracy of the main task and the red line represents the accuracy of the backdoor task.}
    \label{novemnist_exp}
\end{figure*}

Additionally, we compare the performance of NoV with 30 clients but contains different number of attackers, the results are shown in figure \ref{novcifar10_exp} and \ref{novemnist_exp}.
The figure shows that NoV performs well with no more than 20\% attackers within the clients.

\subsection{End-to-End Performance} 

We test the end-to-end performance of NoV on different models and tasks with 30 clients, the threshold of malicious clients is set to 6. 
Table \ref{overhead} shows the computation time and bandwidth in each round. 
The extra computation time is when a malicious Byzantine attacker appears during the aggregation process, the server kicks them out via step x and re-executes step 2. 

\begin{table}[t]
\centering
\caption{Time and bandwidth overhead of NoV on different models and tasks.}
\label{overhead}
\begin{tabular}{cccc}

\thead{\normalsize Models\\ \normalsize and Tasks} & \thead{CNN-\\EMNIST\\(22k params)} & \thead{LeNet5-\\CIFAR10\\(62k params)} & \thead{ResNet20-\\CIFAR10\\(272k params)} \\ \hline

\thead{\normalsize Computation Time\\(s, per round)} & 121.7 & 276.8 & 1334.8 \\ \hline

\thead{\normalsize Bandwidth\\(MB, per round)} & 46.7 & 126.7 & 557.0 \\ \hline

\thead{\normalsize Extra Computation Time\\ \normalsize for Byzantine attacker\\(s, per round)} & 3.0 & 8.9 & 39.7 \\ \hline

\end{tabular}
\end{table}

\section{Conclusion\label{Con}}
Although federated learning can protect data privacy to some extent, it faces many security challenges and threats. 
These threats not only lead to information leakage, but also undermine the performance of federated learning models. 
How to defend against poisoning attacks, privacy attacks and Byzantine attacks simultaneously is a major challenge for secure federated learning. 
In this paper, we propose NoV, a scheme that can verify whether the local model submitted is poisoned while protecting the model privacy, avoiding poison from being aggregated. 
NoV can also kick Byzantine attackers that do not follow the FL procedure out to ensure the execution of the federated learning system with a small time overhead. 
Experiments show that NoV outperforms existing schemes in terms of defense performance. 

\bibliographystyle{ACM-Reference-Format}
\bibliography{main}

\newpage
\appendix

\section{Pseudo-codes of Algorithm\label{pseudo}}
Here we provide the pseudo-codes for the algorithms for generating and reconstructing secret shares, and proving the correctness of decryption.

    \begin{algorithm}[ht]
	\caption{{\rm \textbf{VSS\_Gen (.)}}: Generate verifiable secret shares. \label{vss_generate}}
        \begin{flushleft}
	{\bf Input:}
	Generator $g$, $h$, secret $s$, hiding element $o$, party number $n$, thershold $t$.\\
	{\bf Output:}
	Shares $\{(s_i, o_i)\}_{i\in [1,n]}$, commitments $\{E_i\}_{i\in [0,t-1]}$
        \end{flushleft}
	\begin{algorithmic}[1]
            \State Generate a commitment $E_0=E(s,o)=g^{s}h^{o}$
		\State Randomly choose two polynomial $F, G \in \mathbf{Z}_q[x]$ of degree at most $t-1$ such that $F(x) = s+F_1x+...+F_{t-1}x^{t-1}$ and $G(x) = o+G_1x+...+G_{t-1}r^{t-1}$. 
            \State Commit to $F_i$ and $G_i$ as $E_i=E(F_i, G_i)=g^{F_i}h^{G_i}$ for $i=1,...,t-1$
            \State Compute $s_i=F(i)$ and $o_i=G(i)$ for $i=1,...,n$
            \\
		\Return shares $\{(s_i, o_i)\}_{i\in [1,n]}$, commitments $\{E_i\}_{i\in [0,t-1]}$
	\end{algorithmic}
    \end{algorithm}
    \begin{algorithm}[H]
	\caption{{\rm \textbf{Prv\_Dec (.)}}: Prove correct decryption. \label{prove_decrypt}}
        \begin{flushleft}
	{\bf Input:}
	Public key $pk=(g, h)$, private key $x=\log_g h$, plaintext $m$, ciphertext $c=(c_1, c_2)$.\\
	{\bf Output:}
	Proof $\pi$.
        \end{flushleft}
	\begin{algorithmic}[1]
		\State Randomly choose $a \in Z_q$, compute $A=g^a$, $B=(c_2m^{-1})^a$
            \State Compute challenge $e=Hash(g||h||m||c||A||B)$
            \State Compute $z=xe+a$
            \\
		\Return $\pi=(m,A,B,z)$
	\end{algorithmic}
    \end{algorithm}
    \begin{algorithm}[H]
	\caption{{\rm \textbf{VSS\_Rec (.)}}: Recover the secret. \label{vss_recover}}
	\begin{flushleft}
        {\bf Input:}
	Threshold $t$, $t$ shares $\{(s_i, o_i)\}_{i\in [1,t]}$.\\
	{\bf Output:}
	Secret $S$, $R$.
        \end{flushleft}
	\begin{algorithmic}[1]
            \State Compute $S = \Sigma_{i\in [1,t]} a_i s_i$ and $R = \Sigma_{i\in [1,t]} a_i o_i$ where $a_i=\Pi_{j\in[1,t], j\neq i} \frac{j}{j-i}$
            \\
		\Return secret $S$, $O$
	\end{algorithmic}
    \end{algorithm}








\end{document}